\documentclass{aa}
\usepackage[]{natbib}
\usepackage{graphicx,color}
\usepackage{txfonts}

\begin{document} 

\title{NGC~4104: a shell galaxy in a forming fossil group}

\author{G.~B.~Lima Neto\inst{1}
\and F.~Durret\inst{2}
\and T.~F.~Lagan\'{a}\inst{3}
\and R.~E.~G.~Machado\inst{4} 
\and N.~Martinet\inst{5}
\and J.-C.~Cuillandre\inst{6}
\and C. Adami\inst{5}}

\institute{Instituto de Astronomia, Geof\'isica e Ci\^encias Atmosf\'ericas, Universidade de S\~ao Paulo, Rua do Mat\~ao 1226, S\~ao Paulo, Brazil\\
\email{gastao@astro.iag.usp.br}
\and Sorbonne Universit\'e, CNRS, UMR~7095, Institut d'Astrophysique de Paris, 98bis Bd Arago, 75014, Paris, France
\and N\'ucleo de Astrof\'\i sica, Universidade Cruzeiro do Sul / Universidade Cidade de S\~ao Paulo, R.~Galv\~ao Bueno 868, Liberdade, S\~ao Paulo, SP, 01506-000, Brazil
\and Departamento Acad\^emico de F\'isica, Universidade Tecnol\'ogica Federal do Paran\'a, Rua Sete de Setembro 3165, Curitiba, Brazil
\and Aix Marseille Univ, CNRS, CNES, LAM, Marseille, France
\and IRFU, CEA, Universit\'e Paris-Saclay, Universit\'e Paris Diderot, AIM, Sorbonne Paris Cit\'e, CEA, CNRS,
Observatoire de Paris, PSL Research University, F-91191 Gif-sur-Yvette Cedex, France
}

\date{Received ?? ?? ????; accepted ?? ?? ????}

\abstract
{Groups are the most common association of galaxies in the Universe, found in different configuration states such as loose, compact and fossil groups.}
{We have studied the galaxy group MKW~4s, dominated by the giant early-type galaxy NGC~4104 at $z=0.0282$. Our aim was to understand the evolutionary stage of this group and to place it within the framework of the standard $\Lambda$CDM cosmological scenario.}
{We have obtained deep optical data with CFHT/Megacam ($g$ and $r$ bands) and we have applied both the \textsc{galfit} 2D image fitting program and the IRAF/\textsc{ellipse} 1D radial method to model the brightest group galaxy (BGG) and its extended stellar envelope. We have also analysed publicly available XMM-\textit{Newton} and \textit{Chandra} X-ray data. From $N$-body simulations of dry-mergers with different mass ratios of the infalling galaxy, we could constrain the dynamical stage of this system.}
{Our results show a stellar shell system feature in NGC~4104 and an extended envelope that was reproduced by our numerical simulations of a collision with a satellite galaxy about 4--6\,Gyr ago. The initial pair of galaxies had a mass ratio of at least 1:3. Taking into account the stellar envelope contribution to the total $r$ band magnitude and the X-ray luminosity, MKW~4s falls into the category of a fossil group.}
{Our results show that we are witnessing a rare case of a shell elliptical galaxy in a forming fossil group.}


\titlerunning{NGC~4104: a shell galaxy in a forming fossil group}
\authorrunning{Lima Neto et al.}
\maketitle

\section{Introduction}

According to our current understanding of a cold dark matter and dark energy dominated universe, large scale structure builds hierarchically. There is a gradual assembly of mass in a process that we observe in many different scales \citep[e.g.,][]{Diemand11}. At the super-cluster scale, gravitational collapse is an on-going process, still in the linear phase \citep{Dunner06, Omill15}.

On the other hand, smaller structures such as galaxies and groups of galaxies are already collapsed structures, most of them in quasi-equilibrium but still accreting mass at a very low rate. In our own Galaxy we observe the stellar streams that are probably witnesses of past and current accretion \citep{Helmi99}.

In groups of galaxies, the typical velocity dispersion is of the order of the stellar velocity dispersion inside the galaxy members. Therefore, collisions followed by mergers are one of the most important phenomena that drive galaxy evolution in these environments. These often build up through merger events, leaving morphological features such as tidal streams, stellar shells, rings, and plumes. One evidence for galactic cannibalism is the feature known as shells or ripples in early-type galaxies \citep{Malin80, Quinn84, Athanassoula85}. The first comprehensive catalogue of shell galaxies was done by \citet{Malin83}, who identified 137 galaxies with shells. Shells are concentric interleaved ripples that appear on both sides of the galaxy centre extending to large galactocentric distances.

Although some spiral galaxies reveal the presence of a low surface brightness stellar shell \citep[e.g.,][]{Martinez10,deBlok14}, shells appear to be more common in red early-type galaxies than in blue galaxies \citep[14\%  against  6\% according to][]{Atkinson13},  and  they  are therefore more commonly detected in massive early-type galaxies \citep[galaxies with stellar mass greater than $10^{10.5} M_{\odot}$, cf.,][]{Malin83,Schweizer85,Tal09}. 

Several  mechanisms  for  the  origin  of  shells  have  been proposed over the last decades. The most widely accepted formation scenario advocates that shells are tidal features derived from the result of a minor merger of a smaller galaxy with an elliptical galaxy \citep[see, e.g., recent studies by][]{Amorisco15,Hendel15}, resulting in a series of faint concentric ripples in surface brightness observed throughout the main stellar component. 

Extensive analytical and numerical work supports the merger scenario, in which shells are the  outcome  of  the  process of hierarchical assembly, and they have also been seen in the context of $\Lambda$CDM cosmological simulations \citep[see, e.g.,][]{Cooper10,Cooper11}.  Out of the most massive galaxies in the Illustris simulation, 39 galaxies in a sample of 220 massive ellipticals exhibit shells \citep[][]{Pop2018}. These simulations also confirm that shells in massive galaxies form through mergers with massive satellites. More specifically, mergers with stellar mass ratios $\mu_{\rm star} > 1:10$.

\begin{figure} 
\includegraphics[width=\columnwidth]{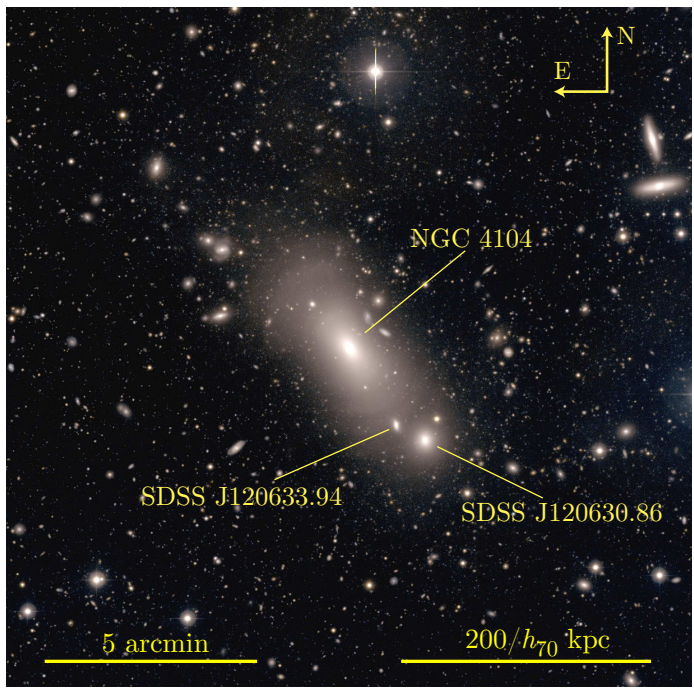}
\caption[]{Centre of the MKW~4s group, showing NGC~4104 and two of the main galaxies which are on the line-of-sight of the extended stellar envelope. This is a composite CFHT/Megacam image which combines two bands, $g$ and $r$ (see text), with size 16$\times$16~arcmin$^2$ ($544 \times 544 h_{70}^{-2}$~kpc$^2$) and is scaled logarithmically in colour. We can see the distinctive boxiness of the envelope. The very faint shells, appearing like ripples along the major axis of NGC~4104, are not easily seen in this image but will be identified below (Sect.~\ref{sec:2Dfit} and Fig.~\ref{fig:NGC4104.r.galfit}).}
\label{fig:NGC4104_10arcmin}
\end{figure}

Any material falling into and passing through a galaxy will be spending more time in the outer region than in the centre (since its velocity will be higher at the centre). Therefore, the outer regions of galaxies should be the locus for assembly clues. Tidal features, and among them shell-like structures, provide a powerful tool to study both the structure and accretion histories of galaxies  \citep{Martinez12,Romanowsky12, Foster14, Amorisco15, Longobardi15}.
For instance, the number  and  distribution  of  shells could be considered to constrain the mass
distribution of the host galaxy, as well as the timing of the
merger  event  itself  \citep{Quinn84,Dupraz86,Canalizo07,Duc16}.
Also, the most extreme examples of ongoing assembly are expected to be the brightest cluster/group galaxies, that reside in the centre of clusters/groups and should involve very active merger histories \citep{Ruszkowski09}. One can also study the dynamics of shell galaxies to date the last merging and physical processes that could be related to that event. For instance, \citet{Ebrova2020} used an $N$-body modelling of NGC~4993 (a shell galaxy) in order to estimate a lower limit for the age of the neutron star binary system that may have been responsible for the short gamma-ray burst observed in 2017.

As far as fossil groups are concerned, observations both in X-rays
\citep{Adami18} and at optical wavelengths \citep{Santos07,LaBarbera09,Girardi14} have led to the idea that these objects are the result of a large dynamical activity at high redshift, located in a too poor large scale environment to evolve
into a cluster. However, \cite{Kim18} found that the NGC~1132 fossil group had a
disturbed asymmetrical X-ray profile, suggesting dynamical activity.
This also seems to be the case in the NGC~4104 fossil group that we
are studying here, as derived from the presence of shells around the central
brightest galaxy detected here for the first time. Therefore, the study
of fossil groups showing traces of recent or ongoing dynamical
evolution is important to illustrate the fact that some structures
considered as ``fossil'' may still be fully ``alive''.

Detecting shells remains challenging specially due to their low surface brightness levels. There are a few tens of known shell galaxies, but NGC~4104 is not yet among them (Fig.~\ref{fig:NGC4104_10arcmin}). Located at 12:06:38.9, +28:10:27 (J2000), it is the brightest galaxy of the X-ray emitting group MKW~4s\footnote{Not be confused with galaxy group MKW~4.} \citep{MKW75,Koranyi2002}, showing shell features and an extended stellar halo. This galaxy is at redshift $z=0.02816$ (from SDSS\footnote{Sloan Digital Sky Survey, \texttt{https://www.sdss.org/}} spectroscopy) and has been studied at many wavelengths \citep{Mulchaey96,LinMohr04,ODea08,Quillen08}, and its optical SDSS spectrum shows strong H$\alpha$ and [NII] lines (i.e. unusually strong for an elliptical galaxy), but very weak [OIII] lines, suggesting that there is star forming activity in  NGC~4104, and that if an active galactic nucleus (AGN) is present it must be very weak.

Also shown in Fig.~\ref{fig:NGC4104_10arcmin} are the second brightest galaxy, SDSS J120630.86+280816.0 with SDSS redshift $z=0.02968$, a red elliptical galaxy, and SDSS J120633.94+280837.5 at redshift $z=0.02543$. These galaxies will also both be modelled in our 2D surface brightness fit.

However, none of the previous studies reported the shell-like features of NGC~4104 and its formation history considering that MKW~4s could be classified as a fossil group. The term fossil group was coined by \citet{Ponman1994} for an X-ray luminous group dominated by a bright elliptical galaxy. More precisely, \citet{Jones03} define a group to be fossil when the difference in the $R$-band between the brightest and second brightest galaxies is larger than 2 magnitudes, both galaxies being within half the group virial radius, and the X-ray luminosity being larger than $10^{42} h_{50}^{-2}$~erg~s$^{-1}$. Such objects are viewed as evolved groups, ``fossil'' of a once dynamically active group of galaxies.
To our knowledge, the only fossil group where shells have been detected around the central elliptical galaxy is that of NGC~1132 \citep{AlamoMartinez12}.

Thus, with deep $r$ and $g$ imaging obtained with the 3.6m CFHT and X-ray publicly available data from the XMM-\textit{Newton} and \textit{Chandra} telescopes, our goal is to understand the evolutionary state of this group and to place it within the framework of the $\Lambda$CDM scenario of large-scale structure formation. $N$-body numerical simulations were performed to give support to our analysis, showing that we are witnessing a rare case of a shell galaxy in a forming fossil group.

Our paper proceeds as follows. In Section~\ref{sec:X} we describe the X-ray data and analysis. In Section~\ref{sec:optical}  we describe the optical data and main results. The dynamical analysis, including $N$-body simulations of the substructure origins is explored in Sect.~\ref{sec:dyn}. In Sect.~\ref{sec:discussion} we discuss MKW~4s as an example of an unusual disturbed fossil group, and in  Sect.~\ref{sec:conc} we summarize the findings of this study. We assume a standard $\Lambda$CDM cosmology: at the redshift of MKW~4s, the group luminosity distance is 123~$h_{70}^{-1}$~Mpc, and 1 arcmin corresponds to $34 h_{70}^{-1}$~kpc.

\begin{figure*}[!htb]
\includegraphics[width=\textwidth]{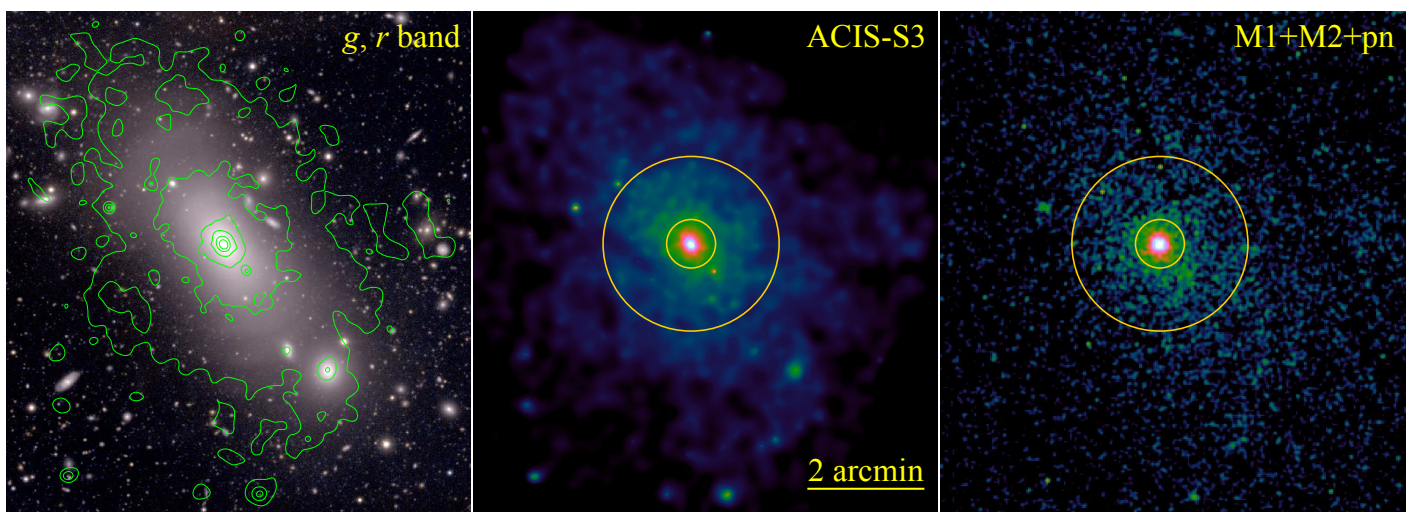}
\caption[]{Left: composite $g$-$r$ image of NGC~4104 with X-ray \textit{Chandra} contours overlaid. Middle: \textit{Chandra} adaptively smoothed image in the [0.5--7.0 keV] band. Right: XMM-\textit{Newton} merged MOS1, MOS2 and pn image in the same energy band as the \textit{Chandra} image. The contours and colours of the X-ray images are in logarithmic scale. The circles on the images, with $R=25^{\prime\prime}$ and $90^{\prime\prime}$, correspond to the extraction regions for the spectral analysis.
The overall shape of the X-ray emission follows the ellipticity of NGC~4104. Notice that the two brightest galaxies on the stellar envelope to the southwest, SDSS~J120633.94 and SDSS~J120630.86, are themselves X-ray sources.}
\label{fig:NGC4104_CFHT_Chandra_XMM}
\end{figure*}

\section{X-ray analysis}
\label{sec:X}

\subsection{X-ray data and data reduction}

The AWM4s group, also called the NGC 4104 group, has been observed in X-rays with various satellites: ROSAT \citep{Dahlem00}, Chandra \citep{Kim19} and XMM-\textit{Newton} \citep{Lagana2013}.

In this section we describe the X-ray analysis. We used \textit{Chandra} (OBSID 3234, P.I. David Buote) and XMM-\textit{Newton} (OBSID 0301900401, P.I. Anna Wolter) publicly available observations that are shown in Fig.~\ref{fig:NGC4104_CFHT_Chandra_XMM}, in comparison with the optical image.

MKW~4s was observed in a single ACIS-S pointing for 30~ks and faint diffuse X-ray emission is detectable to the very outskirts of the ACIS field of view. The \textit{Chandra} data reduction was carried out applying the standard procedure as described in CIAO Threads pages\footnote{http://cxc.harvard.edu/ciao/threads}. The \textit{Chandra} exposure-map corrected image was produced in the broad-band [0.5--7.0 keV] by the script \texttt{merge\_obs} from CIAO.

The  XMM-\textit{Newton} data reduction was done with SAS version 20160201 and calibration files updated in September 2016.  Background flares were identified and rejected by applying a 1.8$\sigma$ clipping to the high-energy count rate histogram.  We used the resulting ``cleaned'' exposure files to detect and exclude point sources from our analysis. 

For both \textit{Chandra} and XMM-\textit{Newton} datasets, the spectral analysis was performed in the [0.5--7.0 keV] band. For the spectral fit, we applied XSPEC v12.9 and $\chi^{2}$ statistics, adopting an absorbed  single temperature plasma model MEKAL \citep{Kaastra93,Liedahl95}, fixing the redshift and letting all the other parameters (temperature, metallicity, normalisation, and hydrogen column density) vary as free parameters. We have adopted the solar metal abundances of \citet{Anders1989}.

\subsection{X-ray results}
\label{sec:xres}
In order to measure the gas mass and the total mass based on hydrostatic equilibrium, as well as the X-ray luminosity, we first need to determine the gas temperature and surface brightness radial profiles.
We start by fitting both \textit{Chandra} and XMM-\textit{Newton} X-ray spectra in a central region ($R < 25^{\prime\prime}$, see Tab.~\ref{tab:Xrayfit25arcsec}), and then in an external annulus, in the region ($25^{\prime\prime} < R < 90^{\prime\prime}$, see Tab.~\ref{tab:Xrayexternalreg}). Both flux and luminosity were computed using \textsc{xspec}, assuming no absorption (i.e., setting $N_{\rm H} = 0$).

\begin{table*}[!htb]
\centering
\caption[]{X-ray spectral fits of the central, $R < 25^{\prime\prime}$ region. Error bars correspond to 90\% confidence level.}
\begin{tabular}{lccccccc}
\hline
        & $T_1$ & $Z_1$ & $T_2$ & $Z_2$ & $f_X$ [0.5--7.0 keV] & $L_X$ (bolom.) & $\chi^2/$d.o.f$^{\strut}$ \\
        & [keV] & $[Z_\odot]$ &[keV] & $[Z_\odot]$ & $[10^{-13}$~erg~s$^{-1}$~cm$^{-2}$] & $[10^{42}$~erg~s$^{-1}$] & \\
\hline
\\[-8pt]
XMM-\textit{Newton}    & $0.73^{+0.09}_{-0.08}$ & 1.0$^\dagger$ & $1.31^{+0.10}_{-0.07}$ & 1.0$^\dagger$ & $3.88^{+0.24}_{-0.26}$ &
           $1.20^{+0.08}_{-0.08}$ & 326.5/247\\[6pt]
\textit{Chandra} & $0.75^{+0.05}_{-0.05}$ & 1.0$^\dagger$ & $1.70^{+0.26}_{-0.21}$ & $0.7^{+0.9}_{-0.5}$ & 
           $5.58^{+0.68}_{-0.17}$ & $1.50^{+0.18}_{-0.04}$ & 113.6/128  \\[6pt]
\hline
\end{tabular}

$^\dagger$ Value held fixed during the fit.
\label{tab:Xrayfit25arcsec}
\end{table*}

\begin{table*}[!htb]
\centering
\caption[]{X-ray spectral fits of the external region, $25^{\prime\prime} < R < 90^{\prime\prime}$ region. Error bars correspond to 90\% confidence level.}
\begin{tabular}{lccccccc}
\hline
        & $T$ & $Z$ &  $f_X$ [0.5--7.0 keV] & $L_X$ (bolom.) & $\chi^2/$d.o.f$^{\strut}$ \\
        & [keV] & $[Z_\odot]$  & $[10^{-13}$~erg~s$^{-1}$~cm$^{-2}$] & $[10^{42}$~erg~s$^{-1}$] & \\
\hline
\\[-8pt]
XMM-\textit{Newton}  & $2.09^{+0.24}_{-0.20}$ & $0.28^{+0.15}_{-0.10}$ & $6.0^{+0.4}_{-0.6}$ & $1.66^{+0.10}_{-0.15}$ & 374.7/361 \\[6pt]
\textit{Chandra} & $2.86^{+0.66}_{-0.51}$ & $0.63^{+0.51}_{-0.31}$ & $5.5^{+0.5}_{-0.4}$ & $1.41^{+0.14}_{-0.10}$ & 208.9/227 \\[6pt]
\hline
\end{tabular}
\label{tab:Xrayexternalreg}
\end{table*}

The central region needed to be fitted by a two-temperature model in order to obtain an acceptable $\chi^2$, i.e., comparable to the number of degrees of freedom of the fit. The fits with only one plasma component yielded 
a reduced-$\chi^2 = 1.6$ and 2.1 for \textit{Chandra} and XMM-\textit{Newton} spectra respectively.
This is probably due to the decrease in temperature in the inner cool-core region which can be mimicked by a two component plasma model.
A cooling flow was indeed reported by \cite{Dahlem00} based on ROSAT data, and a cool-core more recently by \cite{Kim19}, based on \textit{Chandra} data.
These two components are characterized by the indexes ``1'' and ``2'' in Table~\ref{tab:Xrayfit25arcsec}. The temperatures measured by \textit{Chandra} and XMM-\textit{Newton} agree well within the error bars, and combining both fits the resulting mean weighted temperatures are $kT_1 \simeq 0.74$~keV and $kT_2 \simeq 1.5$~keV, for the cooler and hotter components in the inner region. As seen in Table~\ref{tab:Xrayfit25arcsec}, the metallicity was fixed to solar in three of the fits, because the signal to noise ratio was not sufficient to let it free. The metallicity could be left free for \textit{Chandra} component 2, but the large error bars show it is also consistent with being solar.

In the outer region, the mean temperature (using both \textit{Chandra} and XMM-\textit{Newton}) is $kT = 2.30^{+0.45}_{-0.35}$~keV and the mean weighted metallicity is $Z = 0.36^{+0.33}_{-0.21} Z_\odot$. 

The temperature difference between the centre and the outer annulus confirms the previously quoted presence of a cool-core.
Also, the possible difference in metallicity may suggest a gradient, but the error bars are too large for any meaningful conclusion.

Furthermore, we have done a 2D fit of the X-ray intensity map using only the \textit{Chandra} (exposure-corrected, [0.5--7.0~keV] band) image, since it has less artefacts (such as bad pixels and CCD gaps), better signal-to-noise ratio and higher spatial resolution than the XMM-\textit{Newton} image. For this fit we used the \textsc{sherpa} package\footnote{\textsc{sherpa} is part of ciao: \texttt{http://cxc.harvard.edu/sherpa/}} and modelled the diffuse X-ray emission with a $\beta$-model:
\begin{equation}
I(R) = I_0 \left[1 + (R/r_c)^2 \right]^{(0.5 - 3\beta)}\, ;~ R^2 = x^2 + y^2/(1-\varepsilon)^2\, ,
\label{eq:betamodel}
\end{equation}
where $\varepsilon$ is the ellipticity, $(1 - \varepsilon) = b/a$, with $a$ and $b$ the major and minor semi-axes, respectively. The fit is done adding a flat background to the $\beta$-model.
Table~\ref{tab:chandra2Dfit} shows the best fit parameters.

\begin{table}
\caption[]{\textit{Chandra} [0.5--7.0 keV] image best fit elliptical $\beta$-model parameters. Error bars correspond to 68\% confidence level. $\varepsilon$ is the ellipticity.}
\label{tab:chandra2Dfit}
\centering
\begin{tabular}{lcl}
\hline
$r_c$ (arcsec) & $2.66 \pm 0.12$ &($1.50 \pm 0.07$ kpc/$h_{70}$)$^{\strut}$ \\
$\beta$      & $0.54 \pm 0.17$ \\
$\varepsilon$ & $0.25 \pm 0.02$ \\
P.A. (degree) & $31.5 \pm 1.1$ \\
$\chi^2$/d.o.f. & 259849/234242 &= 1.109$_{\strut}$\\
\hline
\end{tabular}
\end{table}

It is not surprising to measure such a small core radius, $r_c=(1.50 \pm 0.07) h^{-1}_{70}$~kpc, which reflects the central cusp in surface brightness, and thus in gas density, associated with cool-core clusters.

\begin{figure}[!htb]
\includegraphics[width=\columnwidth]{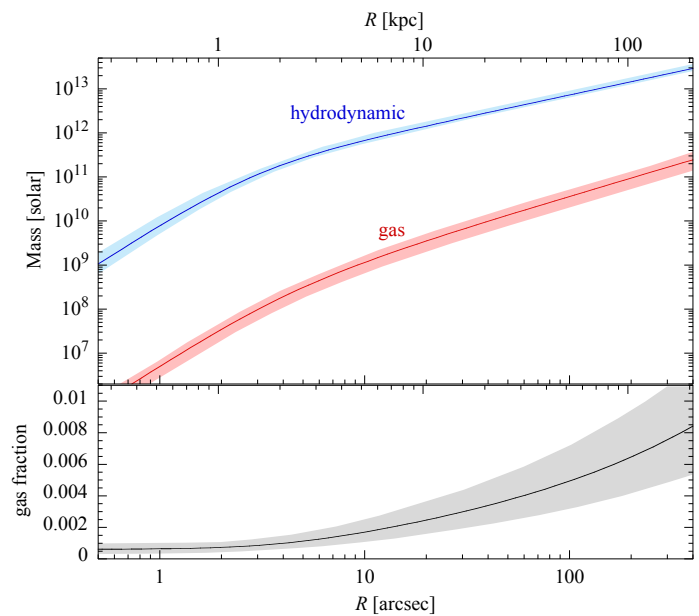}
\caption[]{\textsf{Top}: Hot gas (red) and hydrodynamic (blue) mass profiles derived from X-ray data. \textsf{Bottom}: Gas to total mass ratio. The coloured bands correspond to 1$\sigma$ confidence limits.}
\label{fig:gas_mas_dynamical_fraction2}
\end{figure}

In order to determine the gas density radial profile, we need to calculate the central density of the hot plasma. This is done by normalising the expected emission from the gas distribution (in this case, a simple $\beta$-model) with the so-called emission measure, EM$ = \int n_e n_p dV = 1.18 \int n_p^2 dV$, where $n_e$ and $n_p$ are the electron and proton numerical density, respectively. The numerical factor corresponds to a metallicity $Z = 0.36 Z_\odot$, the mean value obtained by combining XMM-\textit{Newton} and \textit{Chandra} measurements. We have thus normalised the best-fit elliptical $\beta$-model with the EM obtained from the spectral analysis.
The EM is related to the spectrum normalisation used in XSPEC (the factor in front of the integral takes into account the gas metallicity). We considered the emission from an elliptical ring (with ellipticity $\varepsilon = 0.25$) between an inner radius of $25^{\prime\prime}$ and an outer radius of $150^{\prime\prime}$ where we compute the EM. This region is in fact somewhat arbitrary, as far as the surface brightness is well described by a $\beta$-model. Thus, we avoided the inner and outer regions where the surface brightness model may fail.
The central numerical density obtained is $n_0 = (0.33\pm 0.05)$~cm$^{-3}$. With the central density and assuming that $n_e$ follows a $\beta$-model (derived from the deprojection of Eq.~(\ref{eq:betamodel})), we compute the integrated gas mass radial profile, which we show in  Figure~\ref{fig:gas_mas_dynamical_fraction2}.

The total mass is computed by assuming the standard hydrostatic equilibrium hypothesis of the intra-group gas with the group gravitational potential, assuming elliptical symmetry and that the pressure is only thermal \citep[e.g.,][]{Lea1975,Sarazin1986}. We also assume that the intra-group gas is isothermal, with mean temperature derived above ($kT=2.3$~keV).
Figure~\ref{fig:gas_mas_dynamical_fraction2} shows the total mass, baryonic plus dark matter, also known as the hydrodynamical mass radial profile, as well as the X-ray emitting gas fraction profile, $M_{\rm gas}/M_{\rm total}$, for the MKW~4s group.

Based on the total mass radial profile, we can estimate the 
$R_{200}$ and $R_{500}$ radii\footnote{These are the radii where the mean density is 200 and 500 times the critical density of the Universe, respectively.}, and the corresponding masses within these radii:
$$
\begin{array}{ll}
R_{200} = 1075~ \mbox{kpc}~~ (34.6^\prime)\, ; & M_{200} = 1.49 \times 10^{14} M_\odot\, ; \\
R_{500} = ~~680~ \mbox{kpc}~~ (21.8^\prime)\, ; & M_{500} = 0.94 \times 10^{14} M_\odot\, .
\end{array}
$$

In \citet{Lagana2013}, NGC~4104 was fitted with a single temperature using only XMM-\textit{Newton} data, while here we have also used \textit{Chandra} and modelled the X-ray emission as a two-temperature plasma. Also, here we have considered a 2D $\beta$-model to fit the \textit{Chandra} X-ray surface brightness while \citet{Lagana2013} have taken a radial profile based on XMM-\textit{Newton}. The different methodology may explain the different values for $R_{500}$ \citep[in][they found $R_{500}= 502.5$~kpc]{Lagana2013}. This, in turn, results in a different total mass estimation (they have measured $M_{500} = 0.27\times10^{14} M_\odot$).

In order to estimate the total X-ray luminosity up to $R_{200}$ we need to extrapolate the luminosity measured within $R = 90^{\prime\prime}$ (see Tables~\ref{tab:Xrayfit25arcsec} and \ref{tab:Xrayexternalreg}) using the fitted $\beta$-model (Table~\ref{tab:chandra2Dfit}). The X-ray luminosity within $90^{\prime\prime}$, taking into account all the detectors, is $L_X(90^{\prime\prime}) = (2.9 \pm 0.2) \times 10^{42}$~erg~s$^{-1}$.

The projected luminosity as a function of radius is $L_X(R) = 2\pi \int I(R) R dR$, so the extrapolated X-ray luminosity is obtained by integrating Eq.~(\ref{eq:betamodel}), resulting in:
\begin{equation}
L_X^{\rm extrap}(R) = 1.753 [1 - 1/(1 + 0.1413 R^2)^{0.12}] \times L_X^{\rm measured}(90^{\prime\prime}) \, ,
\end{equation}
where we have taken the best fit $\beta$-model parameters. Therefore, we have $L_X(R_{200}) = (4.0 \pm 0.3)\times 10^{42}$~erg~s$^{-1}$.

\begin{figure} 
\includegraphics[width=0.8\columnwidth]{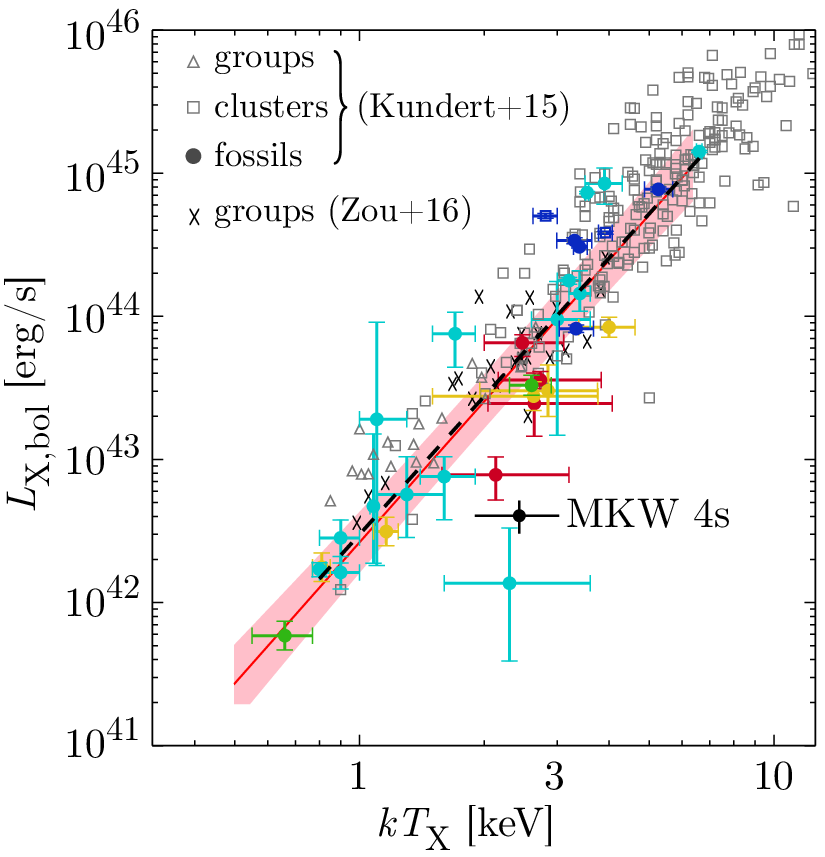}
\caption[]{The $L_X$--$T_X$ relation for a sample of groups and clusters at low and intermediate redshift ($z \la 0.3$) from \citet[][their Fig.~5]{Kundert2015} and \citet[][also their Fig.~5]{Zou2016}. MKW~4s is shown for comparison. The pink shaded area and the red line correspond to the fit by \citeauthor{Zou2016} and the black dashed line is the fit from \citeauthor{Kundert2015}}
\label{fig:LxTx}
\end{figure}

With the above bolometric X-ray luminosity and temperature, MKW~4s is hotter than we would expect from its luminosity when compared to the scaling relation $L_X$--$T_X$ obtained, for instance, by \citet{Kundert2015} (based on a sample of 10 fossil groups observed with Suzaku) and \citet{Zou2016} (based on a sample of 23 \textit{Chandra} observed groups). This appears clearly in Fig.~\ref{fig:LxTx} and may indicate some peculiarity in the MKW~4s system, since fossil groups usually follow the $L_X$--$T_X$ relation defined by non-fossil groups \citep[e.g.,][]{Khosroshahi2007,Kundert2015}. This property may be related with the event that produced the shell structure in MKW~4s (described in detail in Sect.~\ref{sec:2Dfit}), which may have heated the intragroup plasma either from a shock or from feedback of the (possibly) central supermassive back hole. Deeper X-ray observations could resolve this issue. In Fig.~\ref{fig:LxTx}, the blue point at almost the same temperature as MKW~4s but with a lower X-ray luminosity is XMMXCS J030659.8+000824.9, analysed by \citet{Harrison2012}. They note that this system is an outlier of the $L_X$--$T_X$ relation and although it satisfies all criteria to be classified as a fossil group, they reckon this classification is uncertain.


\section{Optical analysis}
\label{sec:optical}

\begin{figure} 
\includegraphics[width=\columnwidth]{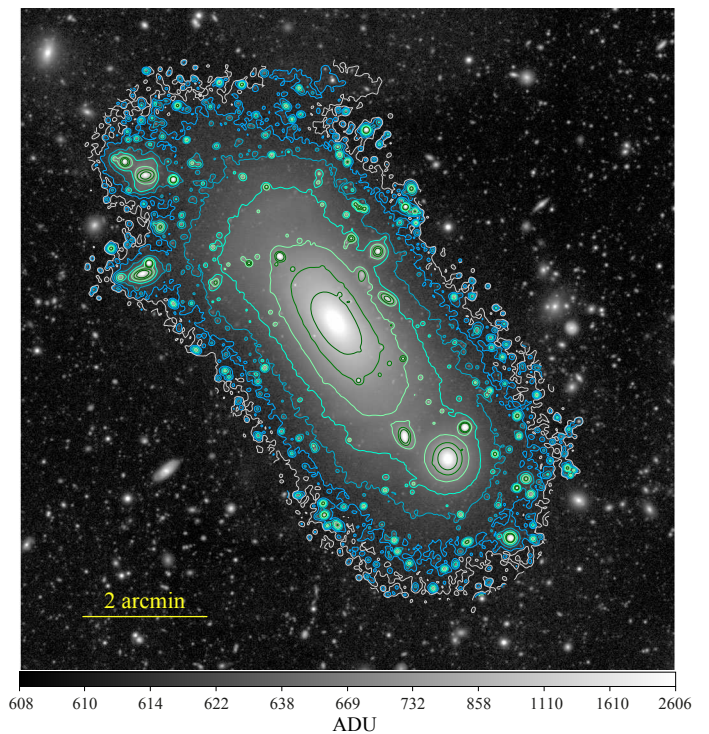}
\caption[]{Grayscale $r$-band image of NGC~4104 with overlaid surface brightness contours spaced from 22 to 28~mag in steps of 1 magnitude (see text for the conversion of ADU to magnitude). The overall shape shows a distinct boxyness. The stellar envelope extends to about 5~arcmin from the centre ($170 h_{70}^{-1}$~kpc) along the major axis.}
\label{fig:NGC4104.r.contGray}
\end{figure}

\subsection{Optical imaging: the data}

We obtained very deep $g$ and $r$ imaging of NGC~4104 with Megacam at the 3.6m Canada-France-Hawaii Telescope (CFHT, observation Program 13AF002 in May and June 2013) and we applied the Elixir-LSB pipeline developed at CFHT \citep{Duc2011}, specially designed for detecting low surface brightness features: dithering during the observations, and then merging the images together. The images we have used were further binned, resulting in a plate scale of $0.561^{\prime\prime}$ per pixel. The exposure time was 2240~s for each band.

The point spread function (PSF) in each band was modelled by a Moffat-function \citep[e.g.,][]{Moffat69, Trujillo01}:
\begin{equation}
\mbox{PSF}(R) = \frac{\beta-1}{\pi\, \alpha} \left[ 1 + (r/\alpha)^2 \right]^{-\beta}\, ,
~\mbox{with FWHM} = \alpha\, 2\sqrt{2^{1/\beta} -1}\, ,
\label{eq:Moffat}
\end{equation}
where FWHM is the full width at half maximum related to the Moffat parameters $\alpha$ and $\beta$.
This allows us to model the surface brightness at radii smaller than the PSF scale (about 1~arcsec), as will be shown in Sect.~\ref{sec:2dfit}.

We choose 30 bright, non-saturated stars around the brightest galaxy, avoiding crowded places and stray light around bright stars.
For each star, we fit a circular Moffat function using the \texttt{IRAF/imexamine} task. For the $r$ band, the best fit values are $\beta = 4.49 \pm 2.13$ and FWHM~$= 1.063^{\prime\prime}\pm 0.071^{\prime\prime}$; for the $g$ band we have $\beta = 4.76 \pm 1.50$ and FWHM~$= 1.065^{\prime\prime}\pm 0.059^{\prime\prime}$.

Furthermore, for the 2D image analysis (Sect.~\ref{sec:2dfit} below), an image of the PSF is produced by stacking non-saturated stellar images after subpixel recentring. The stars are chosen far from the brightest galaxies and far from the refraction stray light around the very bright stars.

Both $g$ and $r$ bands have ZP = 29.61~mag, thus magnitudes are defined as $m = 29.61 - 2.5 \log(\mbox{ADU}/0.3147)$, where the constant 0.3147 corresponds to the plate scale of $0.561^{\prime\prime}$ per pixel.

Using wavelet decomposition\footnote{We have applied the code \texttt{wvdecomp} from the package \textsc{zhtools} developed by A.~Vikhlinin.}, we checked that, for both bands, the images were indeed flat throughout the whole field of view.
The background levels, measured using a 3$\sigma$-clipping method around the median value, are $610.6\pm 2.1$ and $386.8\pm1.8$ ADU per pixel for the $r$ and $g$ bands, respectively. This translates to $21.390 \pm 0.005$ and $21.886 \pm 0.006$~mag/arcsec$^2$ for the $r$ and $g$ bands, respectively. 

Figure~\ref{fig:NGC4104.r.contGray} shows the extension of the stellar envelope down to surface brightness $\mu _r = 28~$mag~arcsec$^{-2}$.  We do not detect intragroup light in the form of streams or substructure, only the smoothly decreasing envelope around NGC~4104. The envelope shows a pronounced boxyness or peanut shape as defined by the 4th cosine coefficient of the Fourier series describing the non-ellipticity of isophotes \citep[e.g., ][]{Lauer1985, Bender1989}.

\subsection{Optical results}
\label{sec:res}

\begin{figure} 
\includegraphics[width=\columnwidth]{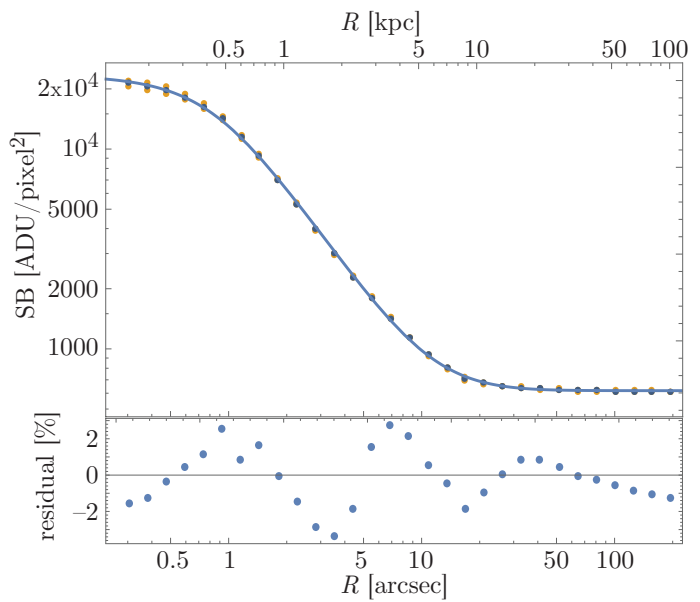}
\caption[]{Best-fit PSF convolved S\'ersic profile for the $r$ band surface brightness profile. The residuals are given in percents relative to the observed profile. The orange points represent $1 \sigma$ error bars.}
\label{fig:ajuste_r_NGC4104_ellipsePSF}
\end{figure}

\begin{figure} 
\includegraphics[width=\columnwidth]{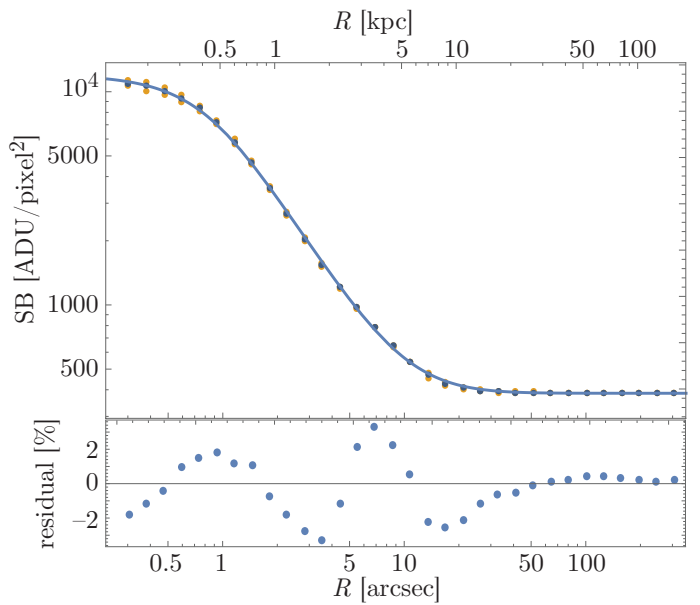}
\caption[]{Same as Fig.~\ref{fig:ajuste_r_NGC4104_ellipsePSF} for the $g$-band.}
\label{fig:ajuste_g_NGC4104_ellipsePSF}
\end{figure}

\begin{table} 
\caption[]{Results of the one component S\'{e}rsic function fit of the radial surface brightness profile obtained with \textsc{ellipse}/IRAF. The ellipticity was held fixed at $\varepsilon = 0.11$ (the mean value determined by the \textsc{ellipse}/IRAF) task, as well as the Moffat parameters, which greatly simplifies the use of Eq.~(\ref{eq:sersicMoffat}). Errors correspond to $90\%$ confidence level.}
\begin{tabular}{lcc}
\hline
Parameter                  &  $r$ band & $g$ band \\
\hline
$R_{\rm eff}$ [arcsec]     & $6.9 \pm 1.6$          & $7.1 \pm 1.9 $\\
$R_{\rm eff}$ [kpc]        & $3.9 \pm 0.9$          & $4.0 \pm 1.1 $\\
$n$                        & $3.29 \pm 0.10$        & $3.41 \pm 0.11 $ \\
$\mu_0$  [mag asec$^{-2}$] & $14.37 \pm 0.02$       & $14.93 \pm 0.3 $ \\
Luminosity  $[10^{10}\, L_\odot]$    & $10.6 \pm 0.3$         & $ 8.4 \pm 0.2$ \\
Total magnitude            & $12.41 \pm 0.07$       & $13.17 \pm 0.08$ \\
backgr. [ADU~pix${}^{-2}$] & $618.0 \pm 6.2$        & $385.2 \pm 3.0$\\
\hline
\end{tabular}
\label{tab:SBellipsefit}
\end{table}

In this section we report the results obtained using two different 2D fitting techniques with the aim of studying the structural components of NGC~4104. For this purpose we have used IRAF/\textsc{ellipse} \citep{Busko1996} and \textsc{galfit} \citep{Peng02} codes, that are briefly described in the sections below.

\subsubsection{Radial surface brightness fit}
\label{sec:2dfit}

For each of the two bands, $g$ and $r$, we followed the same procedure. First, we masked all the bright stars and the stray light coming from diffraction and reflected starlight within the telescope and camera. The diffuse light from stars has a surface brightness comparable to that of the outer galaxy stellar envelope and of the eventual intra-group light.

We have used the \textsc{ellipse} task from IRAF \citep{Jedrzejewski1987,Busko1996}. This task fits the surface brightness with isophote ellipses, without assuming a radial profile model. It applies a 3-sigma clipping technique to eliminate faint point-like emission from the ellipse fitting. A 2D model was then reconstructed with the IRAF/\textsc{bmodel} task. All bright and extended objects in the field were masked to fit the isocontours of NGC~4104. The \textsc{ellipse} task outputs a radial profile along the major axis that we can fit with a parametric model such as the S\'ersic profile.

The radial surface brightness profile that we have adopted is a single S\'ersic profile convolved with a Moffat function describing the point spread function.

\begin{figure*} 
\includegraphics[width=\textwidth]{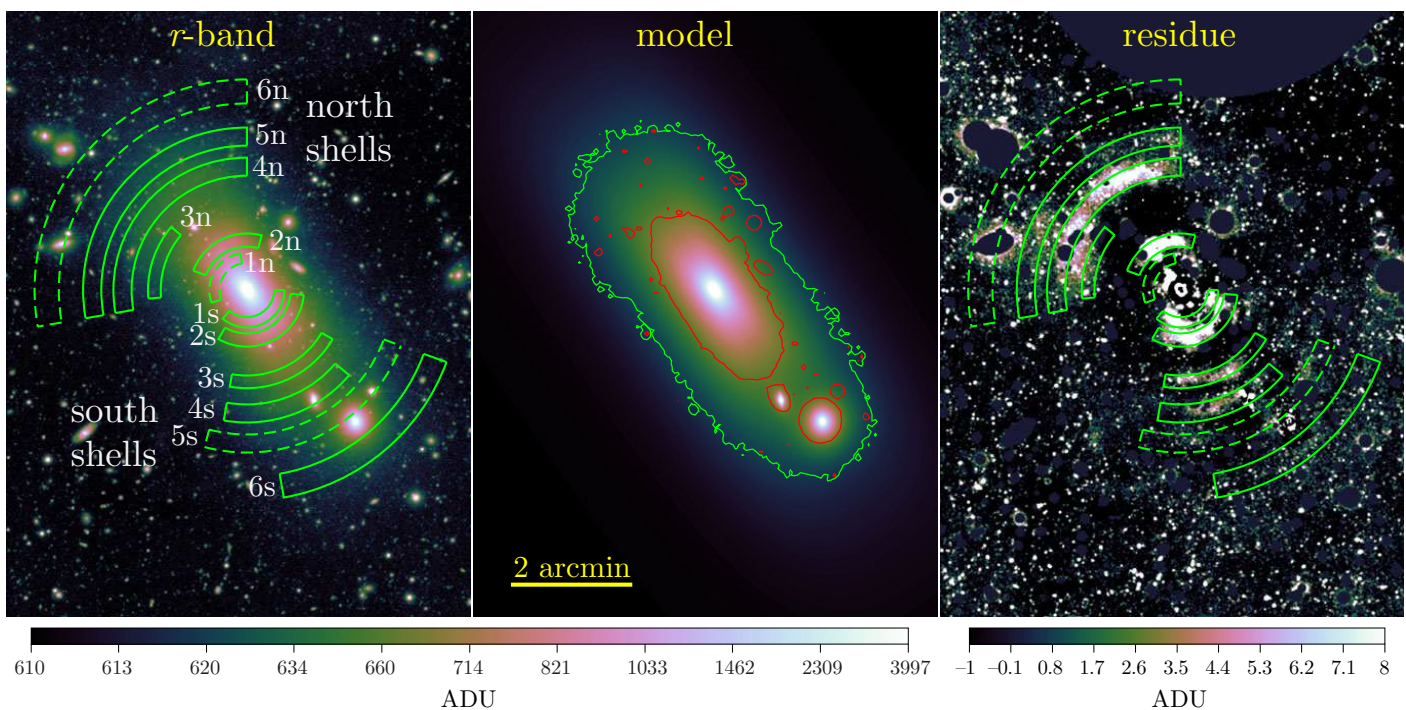}
\caption[]{Best result of the 2D fit using \textsc{galfit}. \textsf{Left}: original $r$-band image with the positions of the various shells labelled from 1 to 6, with ``s'' or ``n'' referring to the southern and northern shells, respectively. The shells were drawn as circular arcs, symmetric with respect to the centre of NGC~4104. The dashed lines correspond to the shells that we could not measure the magnitude, see Table~\ref{tab:shells}. \textsf{Middle}: best fit model (with a flat background included), simultaneously taking into account the three brightest galaxies in the group. The isocontours from the original $r$-band image, corresponding to 24 and 26 mag/arcsec$^2$ (shown in red and green, respectively) are displayed for reference. \textsf{Right}: residue, i.e. original image minus model, with the positions of the shells superimposed. In the left and middle images, the background component is present and the same colour scale is used.}
\label{fig:NGC4104.r.galfit}
\end{figure*}

\begin{table*}[!htb]
\caption[]{\textsc{galfit} 2D best fit results for the $g$ band. Both G1 (NGC~4104) and G3 (SDSS J120633.94) were modelled with two S\'ersic components; G2 (SDSS J120630.86) was modelled with three S\'ersic components (each line corresponds to a different component). The last two columns refer to the sum of the models for each galaxy, not to a single S\'ersic component. The errors are $1 \sigma$ confidence level given by \textsc{galfit}.}
\label{tab:galfit1_g}
\centering
\begin{tabular}{lcrccr|cc}
\hline
\hline
 Obj.  & $g$ mag          &    $R_{\rm eff}$ (kpc)~~ &         $n$           &  $b/a$       &        PA (deg)~~ & $R_{\rm eff}^{\rm tot}$ (kpc) & $L_{\rm tot} (10^{10}L_\odot)$ \\
\hline
G1 & 12.652 $\pm$ 0.002 & 43.12 $\pm$ 0.10 & 1.978 $\pm$ 0.004 & 0.4246 $\pm$ 0.0003 & 34.26 $\pm$ 0.02 \\
G1 & 13.815 $\pm$ 0.005 &  5.25 $\pm$ 0.02 & 1.863 $\pm$ 0.004 & 0.6432 $\pm$ 0.0007 & 29.91 $\pm$ 0.06 & 19.89 $\pm$ 0.09  & 9.45 $\pm$ 0.04 \\
\hline 
G2 & 16.878 $\pm$ 0.024 &  4.07 $\pm$ 0.03 & 0.464 $\pm$ 0.007 & 0.9356 $\pm$ 0.0036 &  7.04 $\pm$ 0.99 \\
G2 & 14.848 $\pm$ 0.005 &  3.16 $\pm$ 0.02 & 4.401 $\pm$ 0.014 & 0.8166 $\pm$ 0.0010 &  5.59 $\pm$ 0.21 \\
G2 & 17.331 $\pm$ 0.042 & 10.38 $\pm$ 0.23 & 0.553 $\pm$ 0.033 & 0.8228 $\pm$ 0.0055 & -62.10$\pm$ 1.88 & 2.74 $\pm$ 0.09 & 1.17 $\pm$ 0.02 \\
\hline
G3 & 17.196 $\pm$ 0.021 &  1.63 $\pm$ 0.01 & 0.971 $\pm$ 0.016 & 0.2966 $\pm$ 0.0019 &  7.79 $\pm$ 0.11 \\
G3 & 16.477 $\pm$ 0.008 &  1.50 $\pm$ 0.01 & 6.031 $\pm$ 0.106 & 0.6103 $\pm$ 0.0030 & 20.84 $\pm$ 0.48 & 1.13 $\pm$ 0.06 & 0.32 $\pm$ 0.01 \\
\hline
\hline
\end{tabular}
\end{table*}

\begin{table*}[!htb]
\caption[]{Same as Table~\ref{tab:galfit1_g} for the $r$ band.}
\label{tab:galfit1_r}
\centering
\begin{tabular}{lcrccr|cc}
\hline
\hline
 Obj.  & $r$ mag          &    $R_{\rm eff}$ (kpc)~~ &         $n$           &  $b/a$              &        PA (deg)~~ & $R_{\rm eff}^{\rm tot}$ (kpc) & $L_{\rm tot} (10^{10}L_\odot)$ \\
\hline
G1 & 12.033 $\pm$ 0.002 & 42.17 $\pm$ 0.09 & 1.740 $\pm$ 0.004 & 0.4202 $\pm$ 0.0003 & 34.25 $\pm$ 0.01 \\
G1 & 12.924 $\pm$ 0.004 &  5.66 $\pm$ 0.02 & 1.914 $\pm$ 0.004 & 0.6262 $\pm$ 0.0005 & 31.06 $\pm$ 0.05 & 17.89 $\pm$ 0.09 & 11.31 $\pm$ 0.05\\
\hline 
G2 & 15.818 $\pm$ 0.049 &  3.98 $\pm$ 0.02 & 0.597 $\pm$ 0.015 & 0.8872 $\pm$ 0.0046 &  3.81 $\pm$ 1.19 \\
G2 & 14.516 $\pm$ 0.071 &  1.86 $\pm$ 0.17 & 3.768 $\pm$ 0.152 & 0.8110 $\pm$ 0.0010 &  9.34 $\pm$ 0.30 \\
G2 & 15.287 $\pm$ 0.115 &  9.20 $\pm$ 0.23 & 1.230 $\pm$ 0.049 & 0.9389 $\pm$ 0.0021 & -22.40$\pm$ 7.37 & 2.60 $\pm$ 0.10 & 1.43 $\pm$ 0.07 \\
\hline 
G3 & 16.567 $\pm$ 0.017 &  1.67 $\pm$ 0.01 & 0.878 $\pm$ 0.012 & 0.2742 $\pm$ 0.0019 &  7.84 $\pm$ 0.08 \\
G3 & 15.662 $\pm$ 0.004 &  1.47 $\pm$ 0.02 & 7.030 $\pm$ 0.119 & 0.6262 $\pm$ 0.0026 & 22.16 $\pm$ 0.42 & 1.14 $\pm$ 0.05 & 0.40 $\pm$ 0.04 \\
\hline
\hline
\end{tabular}
\end{table*}

Taking into account the PSF as a Moffat function, we fitted the following radial profile \citep{Trujillo01}:
\begin{equation}
\begin{array}{l}
\displaystyle I_c(r) = (1- \varepsilon) \int_0^\infty r^\prime I(r^\prime) d r^\prime \int_0^{2\pi} \frac{(\beta -1)}{\pi \alpha^2} \times \\[10pt]
 \displaystyle  \times  \left[ 1 + \frac{r^2 + {r^\prime}^2 - 2r r^\prime \cos \theta^\prime + \varepsilon (\varepsilon -2)(r^\prime \sin \theta^\prime)^2}{\alpha^2} d \theta^\prime
\right]^{-\beta} \, ,
\end{array}
\label{eq:sersicMoffat}
\end{equation}
where we have kept fixed the Moffat parameters, $\beta$ and $\alpha$, using the values shown in Sect.~\ref{sec:optical}, and fixing the ellipticity to the same value for both bands, using the mean value obtained by the \textsc{ellipse} task. For a given radius $r$, the integral in Eq.~(\ref{eq:sersicMoffat}) is computed numerically. Fitting the surface brightness radial profile convolved with the Moffat PSF allows the use of all data points, including the ones with radius smaller than the PSF FWHM, about 1~arcsec for both bands.

The results are shown in Table~\ref{tab:SBellipsefit}, and Figs.~\ref{fig:ajuste_r_NGC4104_ellipsePSF} and \ref{fig:ajuste_g_NGC4104_ellipsePSF}. The best fit S\'ersic parameters are $3.29$ and $3.41$ in the $r$ and $g$ bands respectively, somewhat below the de~Vaucouleurs $n=4$ value. The fits with a single S\'ersic parameter show an oscillating residue (Figs.~\ref{fig:ajuste_r_NGC4104_ellipsePSF} and \ref{fig:ajuste_g_NGC4104_ellipsePSF} lower panels), which may indicate that a more complex model should be considered.

\begin{figure} 
\centering
\includegraphics[width=\columnwidth]{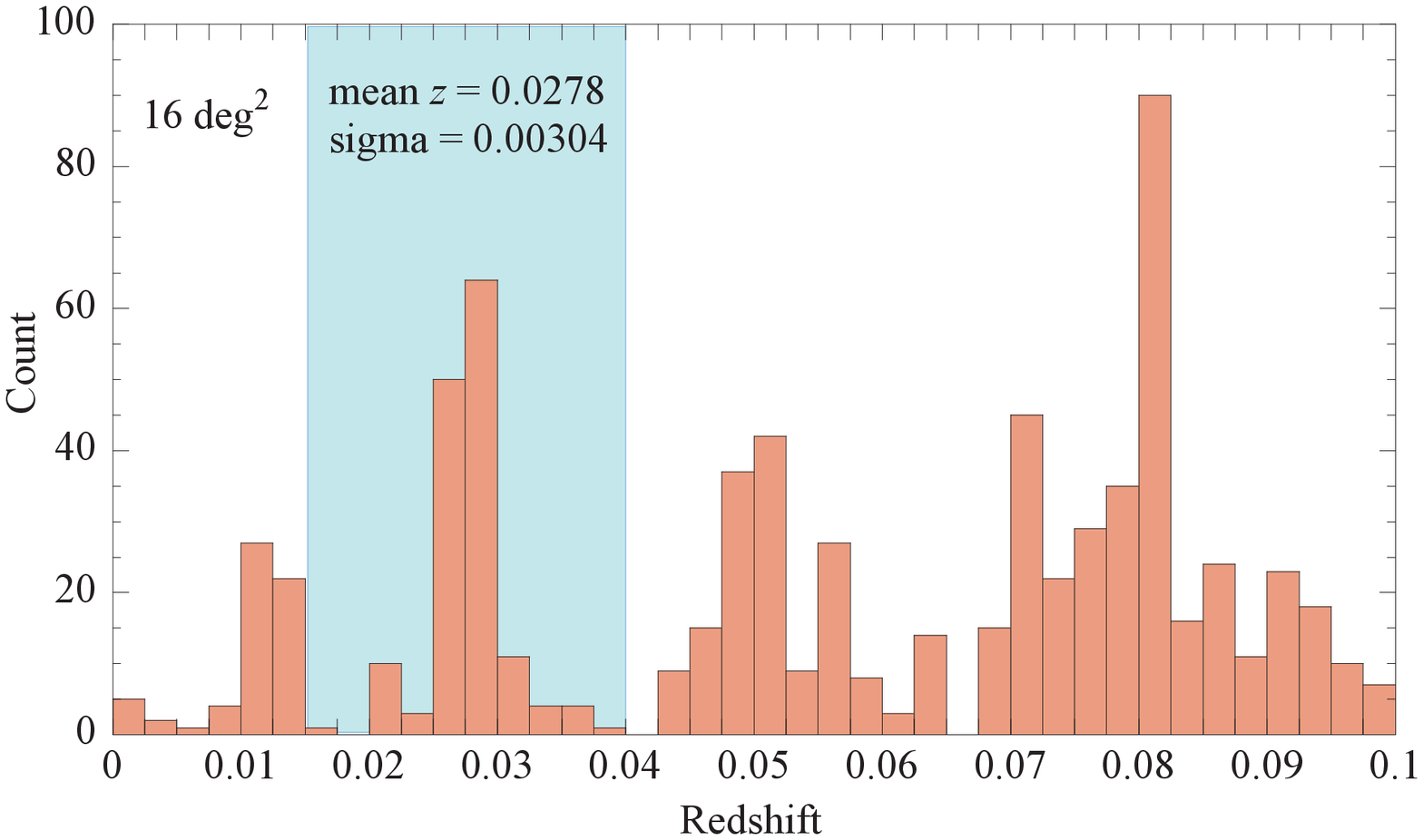}
\caption[]{Redshift histogram of 718 galaxies selected in a $4^\circ \times 4^\circ$ box centred on NGC~4104 with redshift between 0.001 and 0.1 from SDSS. 
There are at least three more structures on the line-of-sight. Focusing on the NGC~4104 structure ($0.015 < z < 0.04$), 
we have $\overline{z} = 0.0278$ and $\sigma_z = 0.00304$ (148 galaxies).}
\label{fig:histogram1}
\end{figure}

\begin{figure} 
\centering
\includegraphics[width=\columnwidth]{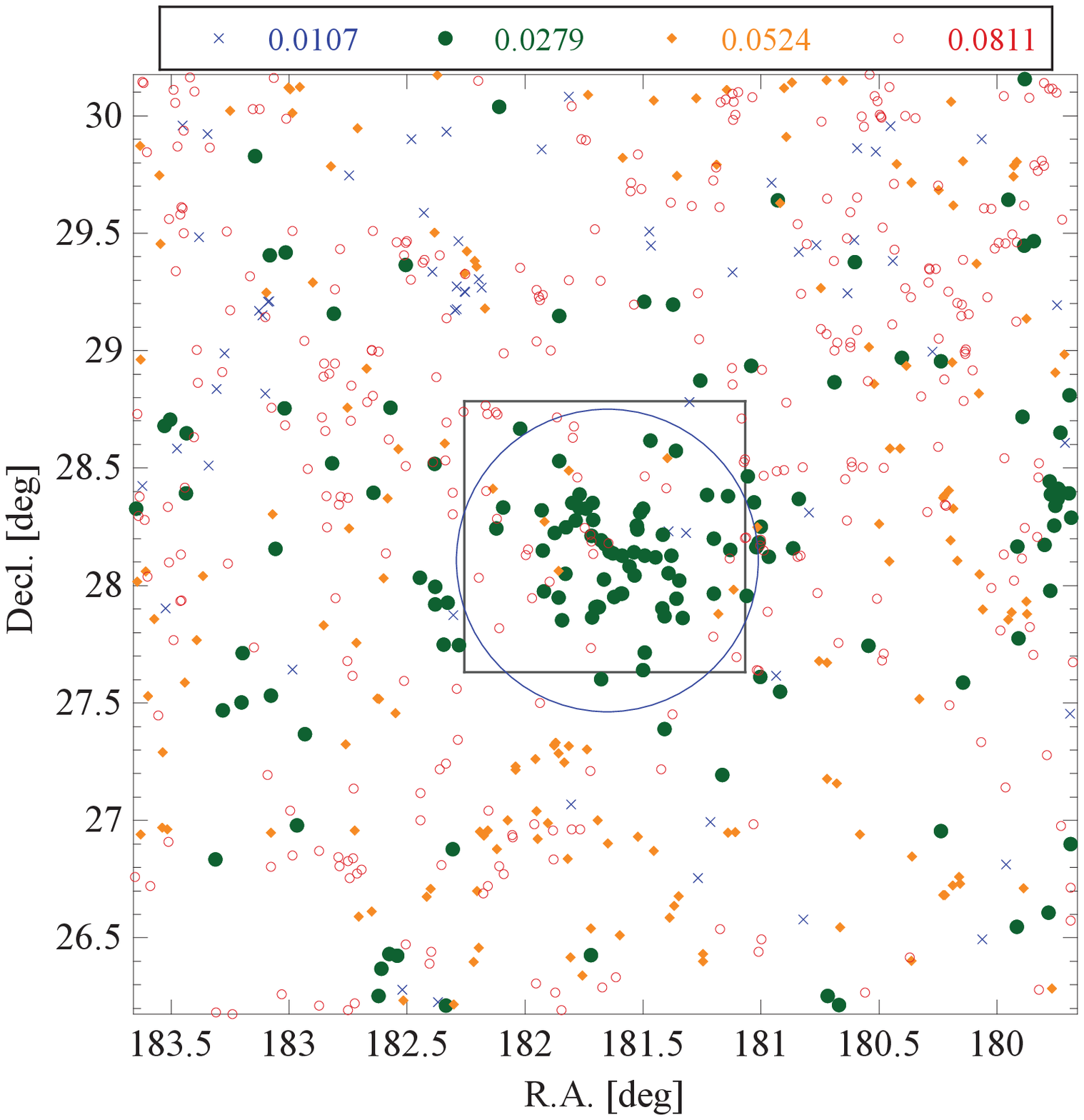}
\caption[]{Spatial distribution of the galaxies shown in Fig.~\ref{fig:histogram1}. The structures identified in the four portions of the histogram are shown with different symbols and colours, identified by their mean redshift just above the figure. The MegaCam field-of-view is shown as a black square. The blue circle has a radius equal to $R_{200}$, as computed in Sect.~\ref{sec:X}, centred on NGC~4104.}
\label{fig:distRADec}
\end{figure}

There are four clear structures 
on the line of sight, as can be seen in the histogram of Fig.~\ref{fig:histogram1}. These four structures are spatially distributed as shown in Fig.~\ref{fig:distRADec}, where each structure is shown with a different colour. NGC~4104 belongs to a structure with 148 galaxies within the redshift range of $0.015 < z < 0.04$ (indicated by the green-filled circles in this figure). The main structure has a redshift $z=0.0278\pm 0.0030$. There is a difference compared with the spectroscopic redshift of NGC~4104, $\Delta z = 0.0004$, but this is smaller than the uncertainty on the mean redshift of the main structure, so we can assume that NGC~4104 is at rest with respect to the MKW~4s group.

\subsubsection{Surface brightness 2D fit}
\label{sec:2Dfit}

With the latest version of \textsc{galfit} \citep{Peng02,Peng10}, it is possible to recover low surface brightness tidal features beneath and beyond luminous galaxies that  can be described by traditional parametric functions such as S\'{e}rsic, Moffat, King etc, profiles. To produce a realistic-looking galaxy model image of NGC~4104, we used a multi-component S\'{e}rsic model: three components that describe the three brightest galaxies (NGC~4104, and the second and third brightest galaxies inside the BGG envelope). All the other objects were masked. The initial conditions were obtained fitting first each galaxy individually, and then fitting the three galaxies simultaneously. The PSF was taken into account by using a synthetic image of a Moffat radial profile described by Eq.~(\ref{eq:Moffat}). The \textsc{galfit} zero point (ZP) is given by $\mbox{ZP}_{\rm gf} = \mbox{ZP} + 2.5 \log(1/T_{\rm exp})$, where $T_{\rm exp}$ is the exposure time.

The best fit parameters are given in Tables~\ref{tab:galfit1_g} and \ref{tab:galfit1_r} for the $g$ and $r$ bands, respectively. In Fig.~\ref{fig:NGC4104.r.galfit} we show the $r$-band image, the best-fit model and the residual image in which the interleaved ripples that appear on both sides of the galaxy centre are evident and coincide spatially with optical stellar shells. 

We have identified the shells, by carefully masking all sources (foreground stars, fore- and back-ground galaxies) using the residual map (Fig.~\ref{fig:NGC4104.r.galfit}), as concentric circular arcs, symmetric with respect to the centre at ($12^{\rm h}06^{\rm m}39^{\rm s}$, $28^\circ 10^\prime 27^{\prime\prime}$) J2000, coincident with the centre of NGC~4104. In Table~\ref{tab:shells} we give their distances from the centre, measured mean surface brightnesses in the $r$-band and mean $g-r$ colours.

Half of the shells (numbers 2, 3, and 4) are detected on opposite sides with respect to the centre of symmetry. Their surface brightnesses range from 24.7~mag/arcsec$^2$ (shell 2n) to 28.9~mag/arcsec$^2$ (shell 6s), both in the $r$-band. They have roughly the same colour index as NGC~4104, but note that the error bars are large.

\begin{table} 
    \centering
    \caption[]{Surface brightnesses in the $r$-band and mean $g-r$ colour indexes of the shells. The shells are symmetric concentric arcs centred on ($12^{\rm h}06^{\rm m}39^{\rm s}$, $28^\circ 10^\prime 27^{\prime\prime}$) J2000. The shell numbers (ID) are the same as in Fig.~\ref{fig:NGC4104.r.galfit}.}
    \setlength{\tabcolsep}{1pt}
    \begin{tabular}{lc|cc|cc}
    \hline
    \hline
            &          &     \multicolumn{2}{c|}{north}  & \multicolumn{2}{c}{south} \\
    \hline
    Shell   &    Radius     &    $\mu_r$  & $g-r$    &     $\mu_r$ & $g-r$ \\
    ID  &   [arcmin]    & [mag/asec$^2$]  & mag &  [mag/arcsec$^2$] & mag \\
    \hline
      1   & $0.5 \pm 0.1$ &         ---     &       ---   & $25.1 \pm 0.3$ & $0.7\pm0.4$ \\
      2   & $0.8 \pm 0.1$ &  $24.7 \pm 0.2$ & $0.8\pm0.3$ & $25.1 \pm 0.2$ & $0.5\pm0.5$ \\
      3   & $1.5 \pm 0.1$ &  $26.7 \pm 0.6$ & $1.3\pm0.6$ & $26.5 \pm 0.4$ & $0.5\pm0.6$ \\
      4   & $2.1 \pm 0.2$ &  $25.5 \pm 0.1$ & $1.0\pm0.8$ & $27.1 \pm 0.4$ & $1.1\pm1.2$ \\
      5   & $2.6 \pm 0.2$ &  $27.1 \pm 0.3$ & $1.2\pm1.1$ &      ---       &   --- \\
      6   & $3.3 \pm 0.2$ &        ---      &       ---   & $28.9 \pm 0.3$ & $0.3\pm0.8$ \\
    \hline
    \hline
    \end{tabular}
    \label{tab:shells}
\end{table}

\begin{figure*} 
\centering
\includegraphics[width=0.85\textwidth]{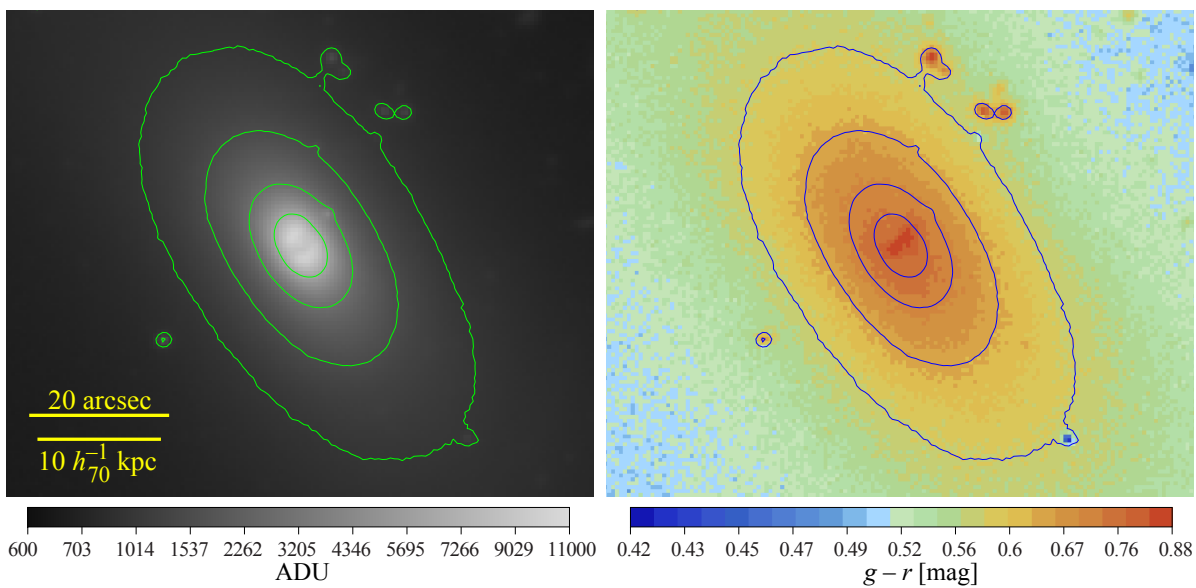}
\caption[]{Left: $r$-band image of the central region of NGC~4104. The contours correspond to surface brightness from 19 to 22 mag~arcsec$^{-2}$ and the yellow bars give the linear scale length. Right: $g-r$ color image of the same region, with the same contours. There is a faint linear structure crossing the image perpendicular to the major axis with a distinctive red colour. Each panel covers an area of $48\times 40 h_{70}^{-2}$~kpc$^2$.}
\label{fig:NGC4104.r.g-rMag}
\end{figure*}

\begin{figure} 
\centering
\includegraphics[width=\columnwidth]{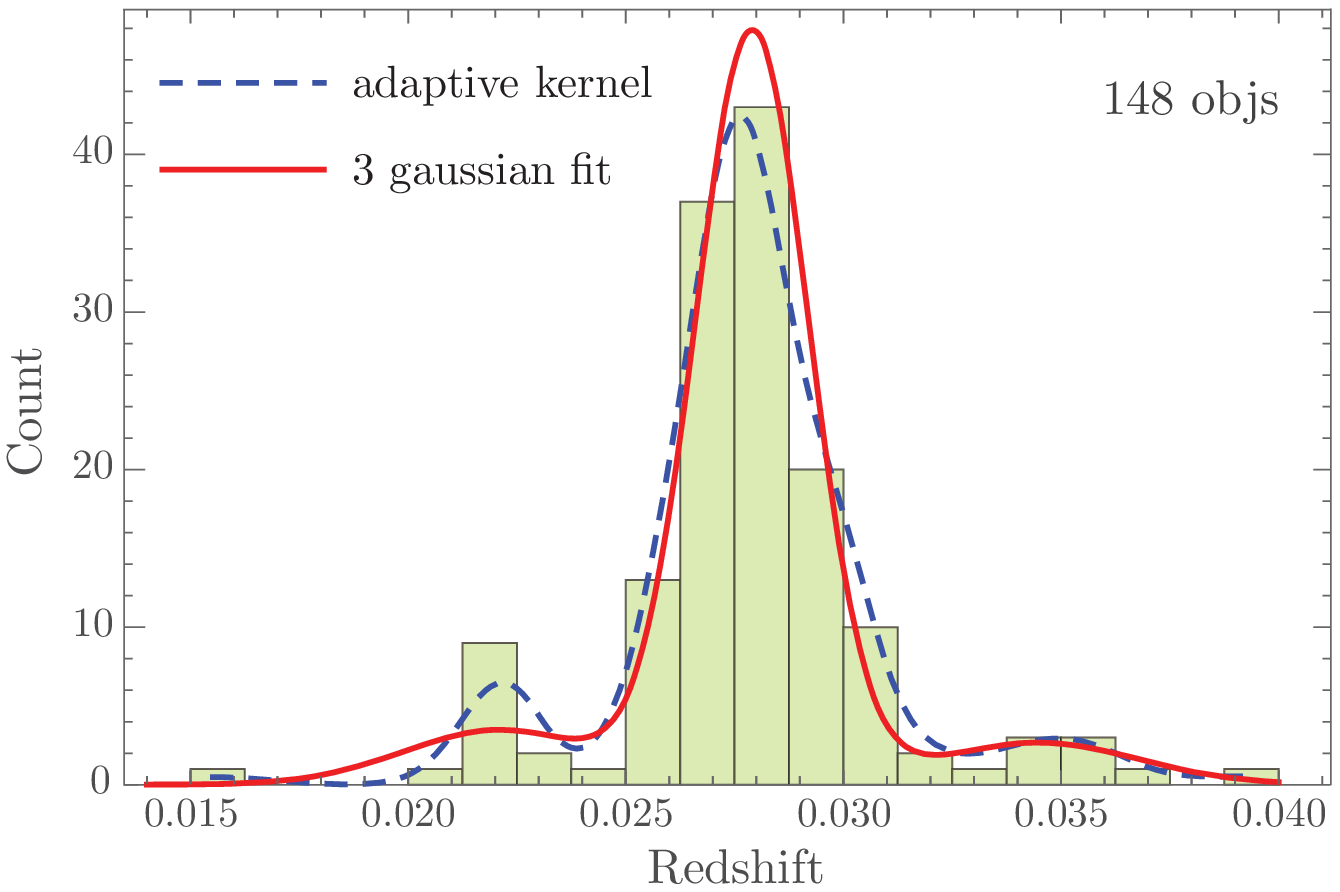}
\caption[]{Redshift histogram of the 148 galaxies of NGC~4104 group. Both a three Gaussian fit and an adaptive kernel 1D distribution suggest a main structure with two smaller substructures in velocity space.}
\label{fig:histo2_zoom}
\end{figure}

\subsubsection{Colour information}

We have measured the colour index in elliptical annuli centred on NGC~4104, excluding the central 2~arcsec (that will be discussed below).

The radial profile of the $g-r$ colour index is very smooth, dropping from $0.70\pm 0.02$ mag in the first annulus (mean major axis of $5^{\prime\prime}$) to $0.50 \pm 0.01$~mag at $\sim 90^{\prime\prime}$ and becoming flat until at least $160^{\prime\prime}$. The shells have the same colour index, as far as we could measure, as the stellar envelope. This would imply that there was no recent or significant star formation when the shells were produced. It then suggests that the galaxy that collided with NGC~4104 was gas poor and it was possibly a dry merger that produced the shells.

\subsubsection{Central region}

Figure~\ref{fig:NGC4104.r.g-rMag} shows the central region of NGC~4104, inside a radius of $24 h_{70}^{-1}$~kpc. On the left panel one can see that there is a filamentary region, perpendicular to the major axis of the galaxy and inside the 19~mag~arcsec$^{-2}$ isocontour, with an approximate elliptical shape with semi-major and minor axes of $2.5^{\prime\prime}$ and $1.0^{\prime\prime}$. 

In the $r$-band, this filament has a surface brightness fainter by about 0.17~mag than the brighter region surrounding it. On the right panel, we can also see that this fainter central filament is also redder, with $(g-r) \approx 0.86$~mag. These images suggest that this filament may be due to an excess of dust in the very inner region of NGC~4104.

This possible dust extinction is also observed on the \textit{Chandra} X-ray image, and suggests the presence of a small central dust lane. This dust may be associated to some kind of activity in the central $1.4 h_{70}^{-1}$~kpc ($2.5^{\prime\prime}$). The evidence, based on the optical spectrum and radio emission \citep{ODea08,Quillen08} is that there is no or a very weak AGN. However, the presence of strong H$_\alpha$ and [NII] lines may indicate that the observed dust in the centre may be related to some recent star formation episode. We may speculate that this event of star formation has been triggered by gas brought by the merging of the galaxy that also produced the shells.


\section{Dynamical analysis of the MKW~4s/NGC~4104 group}
\label{sec:dyn}

\subsection{Galaxy velocity distribution}

We analysed SDSS data (DR12), selecting galaxies in a $4^\circ \times 4^\circ$ box centred on NGC~4104,  with spectroscopic redshifts between 0.001 and 0.1, as shown in Fig.~\ref{fig:histogram1}. There are 718 galaxies in this initial sample.

\begin{figure*}[ht]
\centering
\includegraphics[width=\textwidth]{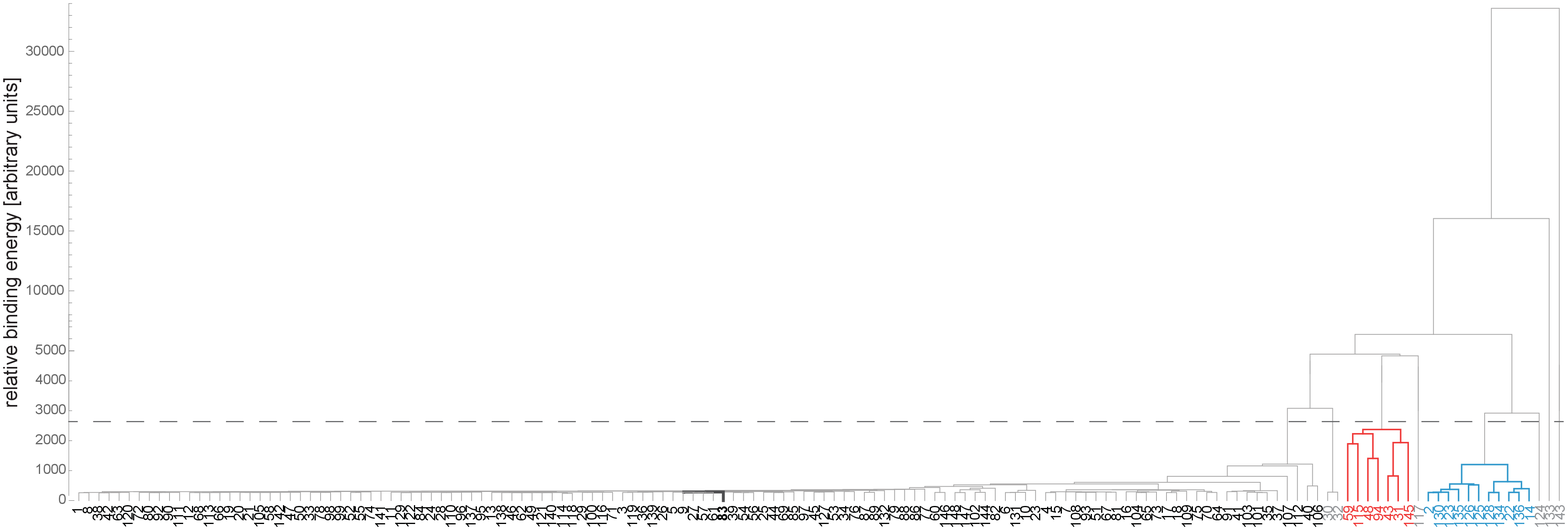}
\caption[]{Dendrogram obtained by applying the \citet{SernaGerbal96} technique for 148 galaxies in the interval $0.015 < z < 0.040$ in a 16~deg$^2$ square centred on NGC~4104 (object number 83 in this figure). The main structure is represented by the black labels (numbers). Two small substructures are shown in red and blue, and a few ``isolated'' galaxies (not strongly bound to any of the three structures) are in grey, on the right side of the horizontal axis. The horizontal dashed  line represents our cut to define the substructures (see text).}
\label{fig:dendro}
\end{figure*}

In Fig.~\ref{fig:histo2_zoom}, we show the redshift histogram of the 148 galaxies of the NGC~4104 group. Both a three Gaussian fit and an adaptive kernel\footnote{We apply the method implemented in \textit{Mathematica}, \texttt{SmoothKernelDistribution} where the histogram is represented by the interpolation of a smooth, continuous function and the kernel varies according to the Silverman (\citeyear{Silverman1986}) rule. The starting kernel that we took had a width of $\delta z = 0.007$ in redshift.} 1D distribution suggest a main structure at $z \simeq 0.0279$ ($\sigma = 0.0013$) with two smaller substructures in velocity space, at $z \simeq 0.0218$ ($\sigma = 0.0021$) and $z \simeq 0.0346$ ($\sigma = 0.0022$). For the main structure, with the highest peak in the histogram, the dispersion in redshift corresponds to a velocity dispersion of 382~km~s$^{-1}$.

\subsection{Substructures around the NGC4104 group}

In order to further analyse the presence of substructures around the NGC~4104 group, we investigate this region by applying the \citet{SernaGerbal96} technique, an agglomerative hierarchical clustering model based on dynamical arguments (pairwise binding energies), to the subsample of 148 galaxies selected in redshift space (as discussed above). This technique has been successfully employed in the past \citep[e.g.,][]{Girardi2011,Barrena2014,Guennou2014} to characterize substructures from a dynamical standpoint. 

The hierarchical clustering algorithm creates a tree of similarities called a dendrogram, that is shown in Fig.~\ref{fig:dendro}. This figure clearly shows three dynamically distinct substructures (two small ones represented in blue and red, and the main one in black), in line with the result obtained from the fit of the histogram of the 148 members. The galaxies marked in blue and red on the dendrogram correspond well to those included in the blue and red features of Fig.~\ref{fig:histo2_zoom}.

This result is consistent with a previous analysis made by \citet{Beers1995} with 53 galaxies, that also suggested the presence of two smaller substructures around NGC~4104. The more prominent of these two substructures can be seen in Fig.~\ref{fig:distRADec}, the green points clustered near coordinates (Ra, Dec) = $(179.5^\circ, 28.4^\circ)$, about 2 degrees to the west of NGC~4104 (about $4 h_{70}^{-1}$~Mpc in the plane of the sky). This West structure agrees with that seen in \cite{Beers1995} who identify it with Zwicky cluster Z1154.9+2806.

\subsection{Large scale structure around the NGC4104 group}

We investigated the isolation degree of the NGC~4104 group of galaxies, as was done in \cite{Adami2007}, section 2, also see \cite{Adami2012}. We search for galaxies brighter than NGC~4104 which may be BCGs of larger groups/clusters of galaxies, that could show that the NGC~4104 group of galaxies is not the dominant one of the considered region. For this, we select two different sky regions, both limited to the blue shaded redshift range in Fig.~\ref{fig:histogram1} (corresponding to a 107 Mpc width at the group redshift). 

The first region is the same as in Fig.~\ref{fig:distRADec}, corresponding to $\sim 8.1$~Mpc on the sky and to a $4^\circ \times 4^\circ$ area. NGC~4104 is clearly the brightest galaxy in this area with an available spectroscopic redshift (the second one is half a magnitude fainter).

We therefore choose the second region, to have a size of the order of typical cosmological voids: $\sim 50$~Mpc (24.5 deg on the sky). In this area, we find in NED three galaxies brighter than NGC~4104 ($g^\prime=13.06$), or fainter than NGC~4104 by less than 0.1 magnitude: NGC~3842 ($g^\prime =12.8$), NGC~4789 ($g^\prime =13.12$), and NGC4555 ($g^\prime =13.1$).

It is thus clear that NGC~4104 is the dominant galaxy of its immediate vicinity (within about 8~Mpc), and is among the dominant galaxies in a volume corresponding to a 50~Mpc diameter. The environment of NGC~4104 is therefore relatively poor in terms of bright galaxies, allowing us to argue that the NGC~4104 group is among the dominant structures of its cosmological bubble, despite its relatively low mass.


\begin{figure}
\includegraphics[width=\columnwidth]{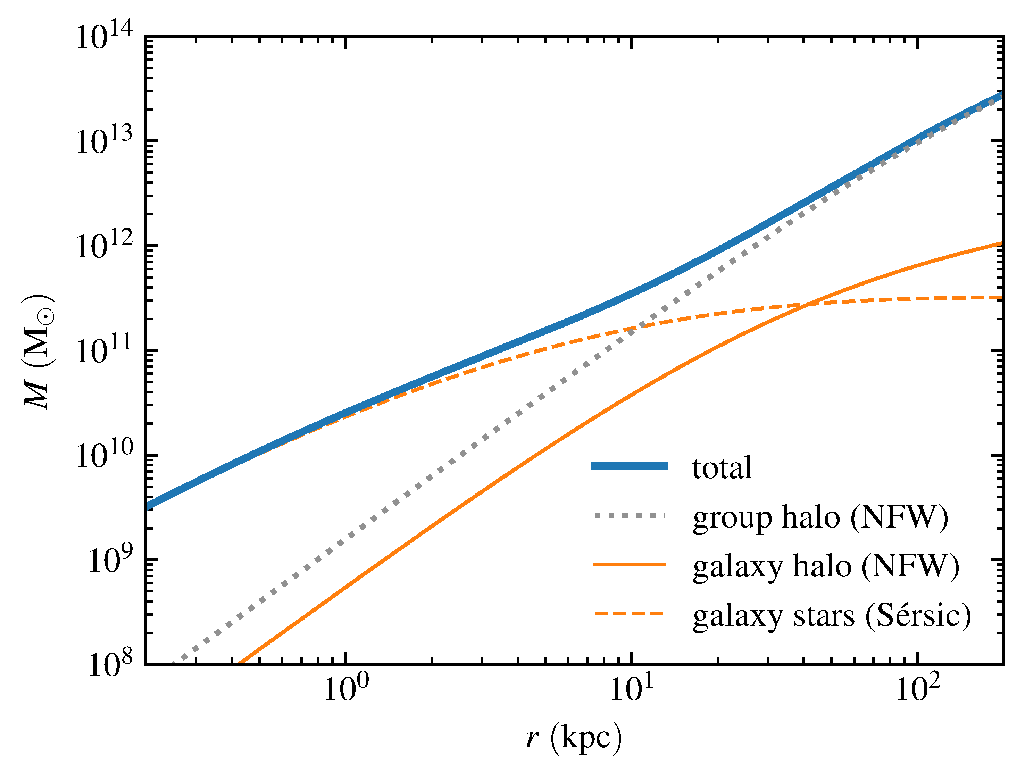}
\caption[]{Cumulative mass profiles of the reference model used to set up initial conditions for the simulations. The elliptical galaxy is composed of stars with a S\'ersic profile (dashed orange line) and a dark matter halo with an NFW profile (solid orange). The galaxy is embedded in a dark matter halo with an NFW profile (grey), corresponding to the group it belongs to.
The total mass profile (blue) is the sum of these three components.}
\label{sim1}
\end{figure}

\subsection{$N$-body simulations}

We have run a series of dry merger $N$-body simulations in order to gain insight on the dynamical process that shaped NGC~4104. Here we present the results of a set of models, noting that this is not meant to be a precise mass reconstruction of the observed system. Rather, these simulations aim to motivate our interpretation of the underlying physical mechanism that led to the formation of the shells, and to broadly attempt to constrain a few of the collision parameters. In this sense, they should be regarded merely as a first-order approximation. We simulated the merger of two galaxies: a more massive galaxy (the ``primary'') and a less massive galaxy (the ``secondary'', not to be confused with the second largest galaxy, J120630.86).

We aim to set up a simulated collision whose outcome is comparable to the observed NGC~4104. The detailed properties of the progenitor galaxies before the collision cannot be known \textit{a priori}. However, the current observed profile of NGC~4104 may be used as a reasonable first guess for the initial conditions. First, we set up a reference mass model inspired by the observational fits for the system. The total mass of the galaxy is obtained from its luminosity (Table~\ref{tab:galfit1_r}) once a mass-to-light ratio is given. Assuming a typical $M/L=10$ for elliptical galaxies \citep{Cappellari2006}, the virial mass of the galaxy would be $M_{200} = 1.1 \times 10^{12}\,M_{\odot}$. Given this mass, and assuming a NFW radial profile \citep{NFW},
the concentration should be approximately $c=7.8$ at low redshift \citep{Duffy2008, Dutton2014}. Consequently, the virial radius is $r_{200}=213$\,kpc and the scale length of the NFW profile is $r_s=27$\,kpc. If a different mass-to-light ratio had been adopted, say $M/L=5$, the virial radius would have been shorter by about 20 per cent. The total mass of the galaxy is dominated by dark matter and to represent the stars, we adopt a S\'ersic profile motivated by the fits from Section~\ref{sec:res}. Thus, a model is obtained for an elliptical galaxy consisting of a dark matter halo described by an NFW profile, and also a stellar component described by a S\'ersic profile. The orange lines in Fig.~\ref{sim1} show the cumulative mass profile of this galaxy, where one sees that the stars dominate only in the inner region but then are overtaken by the dark matter.

The galaxies reside within a group whose total mass was estimated in Section~\ref{sec:xres}. This contribution will also be taken into account, because it affects the collision dynamics at the bottom of the group gravitational potential well. The grey line in Fig.~\ref{sim1} represents the NFW profile of a galaxy group with virial mass $M_{200} = 1.49 \times 10^{14}\,M_{\odot}$, virial radius $r_{200}=1096$\,kpc and concentration $c=4.9$. The NFW scale length is $r_{s}=226$\,kpc. As will be discussed below, for numerical reasons, the profile of the group will be replaced in practice by an equivalent \citet{Hernquist93} profile having a scale length $a=309$\,kpc. In the region of interest the NFW and Hernquist profiles are nearly identical, but the analytical formulae for the Hernquist profile are far more convenient.

The total cumulative mass of the system is then given by the blue line in Fig.~\ref{sim1}. The total enclosed mass within 200\,kpc is nearly $3\times10^{13}\,M_{\odot}$. This is comparable, by construction, to the blue line in Fig.~\ref{fig:gas_mas_dynamical_fraction2}. These mass profiles need not coincide precisely in the innermost region, however. This reference mass model will be used to create the initial conditions of the simulation in the following manner. The mass of the group is fixed. The total mass of the galaxy (stars and dark matter) in the reference model is then divided into the primary and the secondary galaxies, with the mass ratios given in Table~\ref{table1}. This is done such that the total mass of the merged system is always the same. For example, in the simulation labelled S1, the secondary has 10\% of the mass of the primary, and so forth. Thus the two galaxies effectively used in the initial conditions of the simulations are proportionally scale-down versions of the galaxy presented in the reference model of Fig.~\ref{sim1}. The primary galaxy is created in equilibrium in the presence of the group mass. The particles representing the group are then excised, leaving only the bare galaxy. This is done because the group will be replaced by an analytic fixed potential in the calculations. The primary galaxy is then made mildly prolate and begins with a minor-to-major axis ratio of $\sim 0.8$.

To set up the collision, the galaxies were placed at an initial separation of 100\,kpc and with an initial relative velocity of $-100$\,km\,s$^{-1}$ pointed along the major axis of the slightly prolate primary, with impact parameter $b=0$.

The realizations of the initial conditions were performed with the code DICE\footnote{\texttt{https://bitbucket.org/vperret/dice}}\!. The simulations were carried out with the Gadget-2 code \citep{Springel05} with a gravitational 
softening length of 0.1\,kpc, and the evolution was followed for 10\,Gyr. The code was modified to include the forces caused by an analytic Hernquist potential. At each time step of the computations, all dark matter and gas particles feel their own self gravity and additionally the accelerations due to an external fixed potential. In each simulation, the galaxies are represented by approximately $10^6$ dark matter particles and $10^6$ stellar particles, meaning that the mass resolution is better for the stars. With this setting, the mass resolution is $3.5\times 10^5$ and $9.5\times 10^5 M_\odot$, for stars and dark matter particles, respectively.

With the procedures described above, initial conditions were created with the parameters of Table~\ref{table1} and five simulated collisions were carried out with mass ratios from 10 to 90 per cent.

\begin{table*}
\centering
\caption[]{Initial condition parameters of the simulations. Each model has a different mass ratio and the following values are given for the primary galaxy (subscript 1) and the secondary galaxy (subscript 2): virial mass, NFW scale length, concentration and stellar mass.}
\begin{tabular}{cccccccccc}
\hline
\hline
label & $m_2/m_1$  & $M_{200,1}$   & $r_{s,1}$ & $c_1$ & $M_{\star,1}$   & $M_{200,2}$   & $r_{s,2}$ & $c_2$ & $M_{\star,2}$ \\ 
      &            & $(M_{\odot})$ & (kpc)     &       & $(M_{\odot})$   & $(M_{\odot})$ & (kpc)     &       & $(M_{\odot})$ \\
\hline~\\[-0.8em]
S1 & 0.1 & $1.00\times 10^{12}$ & 26.3 & 7.9 & $2.73\times 10^{11}$ & $1.00\times 10^{11}$ & ~9.8 & 9.8 & $2.73\times 10^{10}$ \\
S3 & 0.3 & $8.46\times 10^{11}$ & 24.4 & 8.0 & $2.31\times 10^{11}$ & $2.54\times 10^{11}$ & 14.6 & 9.0 & $6.92\times 10^{10}$ \\
S5 & 0.5 & $7.33\times 10^{11}$ & 23.0 & 8.1 & $2.00\times 10^{11}$ & $3.67\times 10^{11}$ & 17.1 & 8.7 & $1.00\times 10^{11}$ \\
S7 & 0.7 & $6.47\times 10^{11}$ & 21.8 & 8.2 & $1.76\times 10^{11}$ & $4.53\times 10^{11}$ & 18.7 & 8.5 & $1.24\times 10^{11}$ \\
S9 & 0.9 & $5.79\times 10^{11}$ & 20.8 & 8.3 & $1.58\times 10^{11}$ & $5.21\times 10^{11}$ & 19.8 & 8.4 & $1.42\times 10^{11}$ \\
\hline
\end{tabular}
\label{table1}
\end{table*}

\begin{figure*} 
\includegraphics[width=\textwidth]{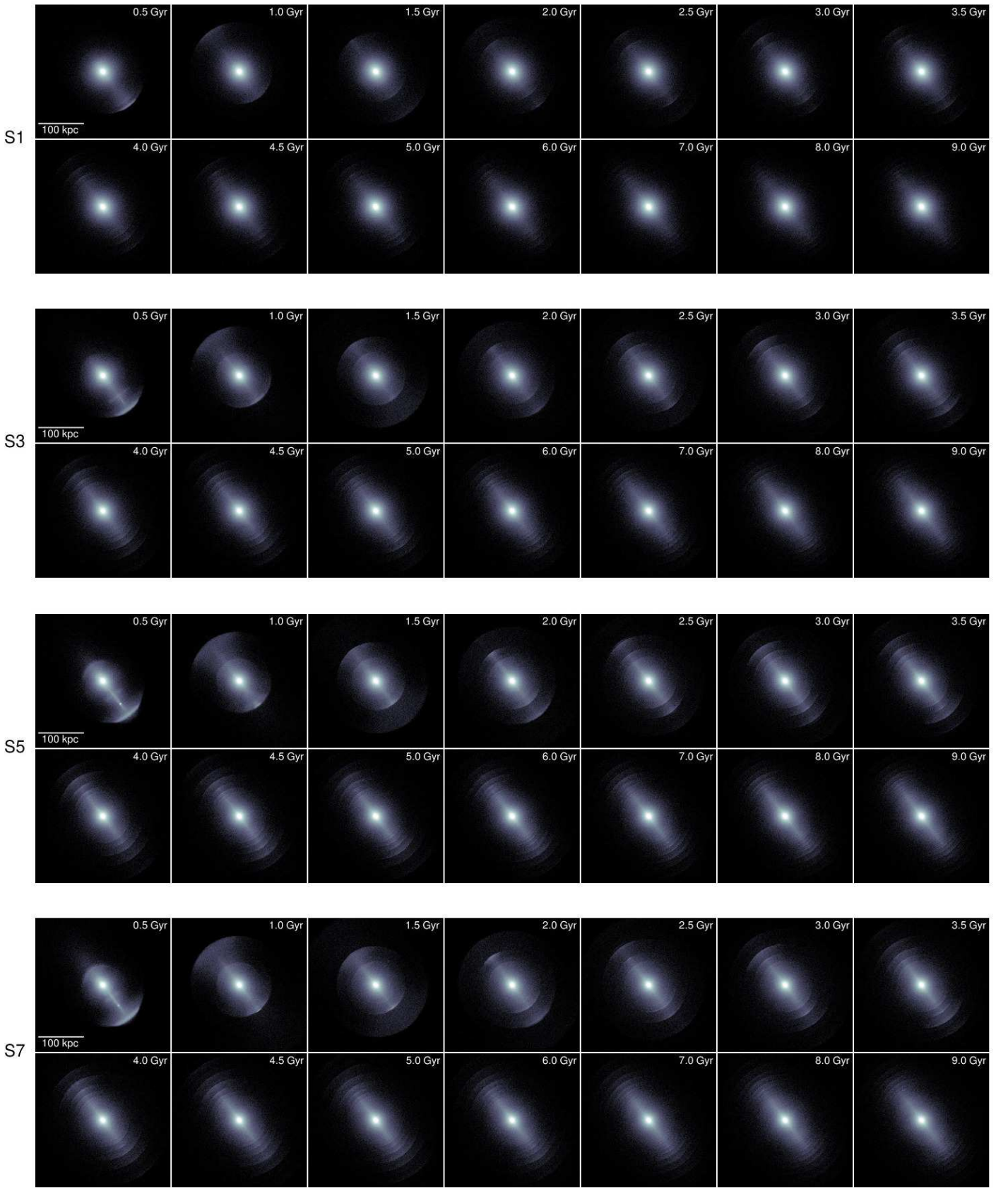}
\caption[]{Time evolution of simulations S1, S3, S5 and S7. Different models correspond to different mass ratios with parameters given in Table~\ref{table1}. Shown are maps of projected stellar mass. Each frame is 300\,kpc wide. Times are given with respect to the instant of central passage. Notice the final frames are more widely spaced in time.}
\label{sim2}
\end{figure*}

In Fig.~\ref{sim2} we present the time evolution of four of these runs, namely simulations S1, S3, S5 and S7. The $300\times300$\,kpc frames display the projected stellar mass during 9\,Gyr, counting from the instant of central passage. The snapshots were rotated merely to match the position angle of NGC~4104, but in Fig.~\ref{sim2} there is no inclination between the collision axis and the plane of the sky. If the aim is to obtain a simulated galaxy whose morphology approximately resembles that of NGC~4104, we found that simulation S1 provided the poorest results. On the other hand, the morphological evolution of models S3--S9 are rather similar to each other (for conciseness, S9 is not shown in Fig.~\ref{sim2}). Comparing the simulations with different mass ratios in Fig.~\ref{sim2}, one finds that throughout their entire evolution, models S3--S7 indeed present some subtle but systematic evolutionary trends. However, their morphological differences are insufficient to confidently rule any of them out. Only model S1 is significantly different, failing to develop the desired shells and overall shape. Thus we conclude that minor mergers of ratio about 1:10 may be ruled out. However, since simulations S3--S9 did not provide a clear constraint, only a lower limit of approximately 1:3 can be offered.

A visual inspection of S3--S7 in Fig.~\ref{sim2} allows us to set an approximate time range for the ``best instant'' of these simulations, i.e. the moments when the morphology appears to be qualitatively similar to the observed NGC~4104. We first notice that the secondary galaxy is disrupted after only a few crossings and within less than a Gyr it is no longer identifiable as a separate object. Also, it is interesting to note that this merging system becomes quite elongated along the collision axis, even though the initial prolateness of the primary galaxy was only mild. Shells develop early on but are noticeably asymmetric at first. Given the symmetry of NGC~4104, we propose that the first few Gyr may be ruled out for this reason -- surely before 2\,Gyr, but possibly before 3\,Gyr. At the other end, shells tend to become more tenuous and the overall shape of the galaxy seems less boxy. These qualitative considerations suggest a preferred time range of 3--8 Gyr.

However, the number of shells might help further constrain this time interval, since the resolution is sufficient to count them. For S3--S7 in Fig.~\ref{sim2}, starting at $\sim 4$\,Gyr, it becomes possible to identify five or six shells on each side. On the other hand, by $\sim 7$\,Gyr seven or eight shells may be counted on each side. These criteria are quite uncertain, because this counting would depend on which of the outermost and innermost shells have sufficient contrast to be discerned; and these qualitative judgements would be based on mass distribution rather than light. It would be doubtful that at a precise instant one would be able to match precisely each observed shell to each simulated one. Thus the approximate time interval of 4--6\,Gyr suggested by counting of shells should be regarded with less confidence than the 3--8\,Gyr range. In any case, within either of these intervals, the maximum extent of the shells reaches 100--150\,kpc (depending on whether the outermost very tenuous shells should be counted), in fair agreement with the observed value.

\begin{figure*} 
\centering
\includegraphics[width=\textwidth]{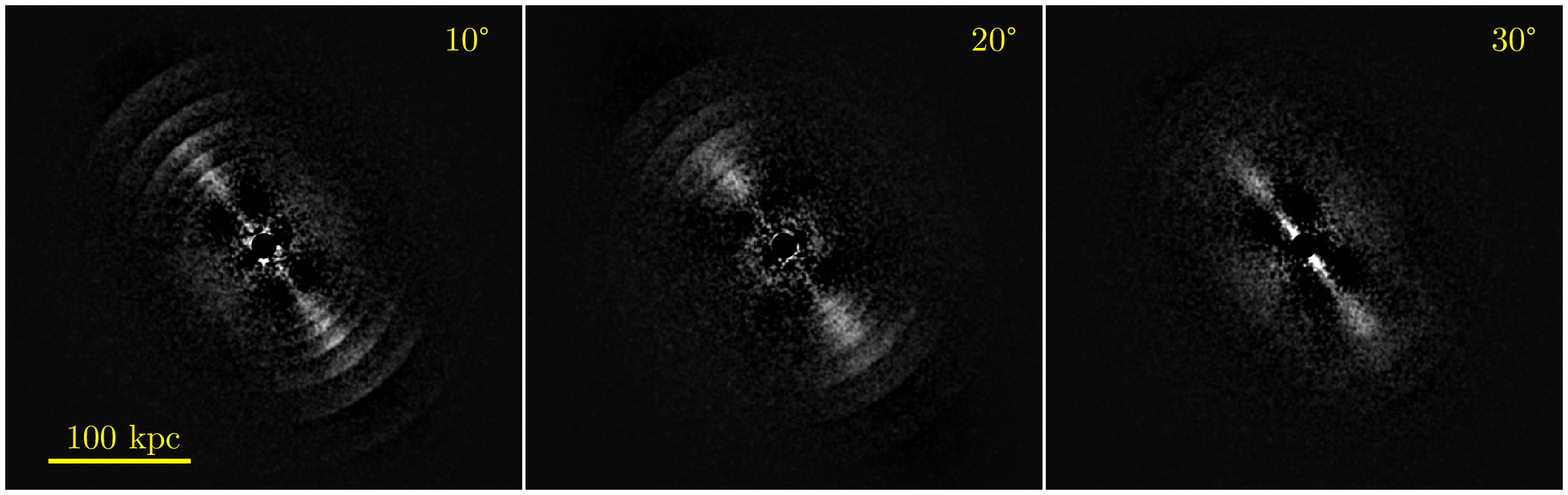}
\caption[]{Residual images computed by subtracting an elliptical symmetric model obtained by fitting a synthetic image \textsc{galfit}, following the same procedure as with NGC~4104 (see Fig.~\ref{fig:NGC4104.r.galfit}). The synthetic image was generated using only the stellar component of simulation S3 at $t=5$\,Gyr. From left to right, we show three different projection angles of the collision axis on the plane of the sky, $10^\circ$, $20^\circ$, and $30^\circ$. The middle panel shows our preferred projection, where the number and intensity of the shells on the residual image are closer to what we observe on NGC~4104.}
\label{fig:simula_residuo}
\end{figure*}

In order to make a more quantitative comparison between the simulations and NGC~4104, we have produced synthetic images by projecting the particles representing the stars onto a plane. Then, this projection was written into a FITS file with the same size and angular scale as the observation. For these images, we have proceeded as we did with NGC~4104, by fitting an elliptical symmetric model with \textsc{galfit}.
Figure~\ref{fig:simula_residuo} shows the residual images after subtracting the \textsc{galfit} best-fit model. We show the images that best approach the shell position and intensity as observed in NGC~4104, which is simulation S3 at $t=5$\,Gyr.

We also explored the possibility that the collision that generated the shells were not exactly on the plane of the sky. Therefore, we also produced synthetic images assuming
projections at $i=10, 20~{\rm and}~30^{\circ}$, where $i$ is the angle between the collision axis and the plane of the sky. The sharpness of the shells in the residuals is sensitive to the inclination. We found that for $i=0$ the simulated shells are much sharper than the observed ones, while for $i\gtrsim30^{\circ}$ they nearly vanish. An angle of about $20^\circ$ seems to best reproduce the observed shells, both in number and intensity.

Shell galaxies have been found to form within cosmological simulations of galaxy formation. In fact, out of the most massive galaxies in the Illustris simulation, nearly 18 per cent exhibit shells \citep{Pop2018}. It is interesting to note that our highly idealised simulations give rise to morphologies resembling those seen in cosmological simulations. For instance, \citet{Torrey15} analyse galaxies from the Illustris simulation, whose synthetic images are not too dissimilar from idealised binary mergers. More generally, the statistical findings of \citet{Pop2018} show that shell galaxies at redshift $z=0$ in Illustris form preferentially from mergers with mass ratios greater than 10 per cent, that progenitors are accreted from orbits of low angular momentum (i.e.~nearly radial) and that the preferred time-window of mergers is 4--6\,Gyr ago. In our attempt to constrain the age of the merger of NGC~4104 and the masses of its progenitors, we have obtained ranges of time and mass which are compatible with the global statistics of cosmological simulations. Even though these estimates are somewhat uncertain, their general consistency would indicate that the essential physical mechanisms are being sufficiently well represented by the simplified simulations, in spite of their many limitations. In other words, the main processes that shaped NGC~4104 must have their origin in stellar dynamics, governed by gravity.

We have probed only one dimension of the parameter space, namely the mass ratios, while keeping all other parameters fixed. The evolution of these simulated collisions has allowed us to estimate rough intervals of age and of mass ratio. The intervals that we obtained are broadly compatible with the statistical predictions from cosmological simulations \citep{Pop2018}. A fuller exploration of the parameter space of possible collisions (and possible initial galaxies) is a large undertaking beyond the scope of the current work. Additional variations of parameters might include different initial velocities, non-frontal collisions, the axis ratios of the initial galaxies, projection effects, and the inclusion of gas and star formation. These are relevant prospects both for tailored simulations that attempt to model individual objects such as NGC~4104, but also for simulations aiming to understand the phenomenon of shell formation in general.

One can never claim that a given solution is unique when attempting to model in detail one particular object. Even though in our specific runs the resulting morphologies exhibit acceptable agreement around 4--6\,Gyr, simulations with other combinations of parameters might have led to other best moments. 
Nevertheless, as a first approximation, they allow us to offer a somewhat qualitative but physically plausible scenario for the merger event that may have given rise to the shells of NGC~4104.

\section{Discussion}
\label{sec:discussion}

While galaxies displaying shells have recently been dynamically disturbed, we have on the other side fossil groups, which are thought to be  old systems that have relaxed after the elliptical galaxy assembly. Fossil groups are X-ray-bright galaxy groups dominated by a large elliptical galaxy and the ``fossil'' nomenclature indicates that such systems are the remnants of galaxy mergers, in which the elliptical galaxy with an X-ray halo is the only result of the process. 

\citet{Jones03} defined fossil groups as spatially extended X-ray sources with $L_X > 5 \times 10^{41}~h_{70}^{-2}$~erg~s$^{-1}$ and with an $R$ band magnitude gap of 2 or more between the brightest and second brightest galaxy member ($\Delta m_{12}$), inside half of the group virial radius. Recent studies of local fossil groups and clusters show that their X-ray luminosities are in the range of $10^{42}-10^{45}$~erg~s$^{-1}$ \citep{Girardi14,Bharadwaj16}, 1--3 orders of magnitude higher than those of typical giant elliptical galaxies with similar optical luminosities \citep[e.g.,][]{Kim15}. Their X-ray surface brightness is usually extended and smooth, as expected for old relaxed systems, and also, the $\Delta m_{12}$ conditions ensure the systems are indeed relaxed and evolved after the elliptical galaxy assembly.

The most widely accepted scenario for the formation of fossil groups is one where dynamical friction causes galaxies close to the centre of the group to merge and a large elliptical galaxy with a hot X-ray halo is left \citep{Donghia05}. However, it is still unclear if the properties of fossil systems make them a particular class of system, and also if they are the final stages of mass assembly or are a transient phase in the formation of larger structures. Based on 25 objects (SDSS and ROSAT All-Sky Survey data), \citet{LaBarbera09} find no difference between fossil groups and non-fossil groups, suggesting that they are the final stages of mass assembly in a region without enough surrounding matter.

For the MKW~4s group, NGC~4104 is 2 mag brighter than the 2nd ranked galaxy in the $r$ band, satisfying the $\Delta m_{12}$ condition. Also, the X-ray luminosity is brighter than $5 \times 10^{41}$~erg~s$^{-1}$, suggesting MKW~4s is a fossil group.

Contradicting the ``fossil'' nomenclature, NGC~4104, which is the brightest galaxy of MKW~4s, shows shell features, that are imprints of a merging, which happened roughly between 3 to 8~Gyr ago (or, possibly between 4 to 6~Gyr ago), with a galaxy that had at least $\sim 30$\% of NGC~4104 mass at that moment. This makes MKW~4s an unusual and disturbed fossil group, as NGC~1132, for which \cite{Kim18} suggested a formation scenario where the growth of the massive central galaxy occurs through continuous infall from its environment; this infall may continue even after the group has entered the fossil phase \citep{Dariush10,Kanagusuku16}. Contrary to NGC~1132 -- and closer to the ``fossil group'' definition --, the X-ray surface brightness of NGC~4104 shows a rather symmetrical morphology, aligned with the optical counterpart, and a cool-core, suggesting that it is a more dynamically evolved system than NGC~1132. To our knowledge, the only fossil groups known until now with shells around their central galaxies are NGC~1132 and NGC~4104. According to  \cite{Aguerri11}, the Fossil Group Origins (FOGO) survey is only expected to reach a limiting surface brightness of $\sim$~26 mag~arcsec$^{-2}$, so it should have detected some shells if there were any, even though the magnitude limit is barely deep enough. Using the same instrument and method as us, \cite{Bilek16} reached 29~mag~arcsec$^{-2}$ in the $g$ band and detected shells around the elliptical galaxy NGC~3923, but this galaxy is not in a fossil group.

\section{Conclusions}
\label{sec:conc}

We have measured the surface brightness radial profile along the major axis using the \textsc{ellipse} task from IRAF. We fitted it with a single S\'ersic profile 
taking into account the PSF, which was modelled by a Moffat function, that allowed us to consider data points (pixels) inside the PSF length scale (of about 1 arcsec). This method, however, cannot fit properly the extended stellar envelope. Therefore, we used \textsc{galfit}, a 2D fitting program, to model NGC~4104 plus the second and third 
brightest galaxies inside the BGG envelope. The residual image clearly shows the shell features that were probably created by the merger of the second 
brightest galaxy with NGC~4104, as corroborated by our $N$-body simulation results.

From the X-ray data, MKW~4s could be classified either as a rich group or a poor cluster, with virial temperature $\sim 2.3$~keV and bolometric X-ray luminosity (extrapolated to $R_{200}$) $\sim 4 \times 10^{42}$~erg~s$^{-1}$. 
The X-ray surface brightness morphology follows the optical counterpart, both being elongated with ellipticities of 0.25 and 0.15, for the X-ray and optical components, respectively. The X-ray emission suggests a dynamically evolved system, with a cool-core and no notable asymmetries in the surface brightness emission.

From our dynamical analysis, the NGC~4104 group is certainly formed by three dynamically distinct substructures: two small structures and a main one. This result was obtained both by a clustering technique and by fitting three Gaussians and an adaptive kernel to the redshift histogram.
This indicates that the NGC~4104 group has recently assembled and its history could explain the shell features.

We also carried out a set of $N$-body simulations of mergers. These numerical models suggest that a radial collision with a satellite galaxy may have led to the formation of the observed shell features. Simulations allowed us to rule out minor mergers with mass ratio 1:10 and to set a lower limit of at least 1:3, and an upper limit as high as 90\%, an almost equal mass collision. The age of the collision was estimated to be at least in the range 3--8\,Gyr, but possibly 4--6\,Gyr with lower confidence.

We also note that neither the optical imaging nor the $N$-body simulations show any sign of intra-group light down to magnitude 28 in the $r$ and $g$ bands, except for the extended stellar envelope around NGC~4104. There is no sign of features resembling tidal debris or ``plumes'' of stars such as the ones observed in more massive systems, for instance, in the Virgo cluster \citep{Mihos2017}. In this sense, NCG~4104 would be dynamically older than some of these massive systems or would have had a less eventful dynamical history.

Given the magnitude gap between the first and second brightest galaxies, it seems that we are witnessing the formation of an object that falls within the fossil group classification but still has scars from a past dynamical interaction.

\begin{acknowledgements}
We thank the referee for constructive suggestions that helped improve the clarity of this paper. GBLN is grateful for the financial support from FAPESP (grant 2018/17543-0) and CNPq, and thanks the hospitality of IAP. TFL acknowledges financial support from FAPESP and CNPq (through grants 18/02626-8 and 303278/2015-3, respectively). REGM acknowledges support from CNPq through grants 303426/2018-7 and 406908/2018-4. F.D. acknowledges long-term support from CNES. NM acknowledges support from a fellowship of the Centre National d'Etudes Spatiales (CNES). This work  made use of the Laboratory of Astroinformatics (IAG/USP, NAT/Unicsul), whose purchase was made possible by the Brazilian agency FAPESP (grant 2009/54006-4) and the INCT-A. Based on observations obtained with MegaPrime/MegaCam, a joint project of CFHT and CEA/DAPNIA, at the Canada-France-Hawaii Telescope (CFHT) which is operated by the National Research Council (NRC) of Canada, the Institut National des Sciences de l'Univers of the Centre National de la Recherche Scientifique of France, and the University of Hawaii (observation Program 13AF002). The authors acknowledge the National Laboratory for Scientific Computing (LNCC/MCTI, Brazil) for providing HPC resources of the SDumont supercomputer, which have contributed to the research results reported within this paper. 
\end{acknowledgements}

\bibliographystyle{aa} 
\bibliography{refs} 

\begin{thebibliography}{85}
\expandafter\ifx\csname natexlab\endcsname\relax\def\natexlab#1{#1}\fi

\bibitem[{{Adami} {et~al.}(2018){Adami}, {Giles}, {Koulouridis}, {Pacaud},
  {Caretta}, {Pierre}, {Eckert}, {Ramos-Ceja}, {Gastaldello}, \&
  {Fotopoulou}}]{Adami18}
{Adami}, C., {Giles}, P., {Koulouridis}, E., {et~al.} 2018, \aap, 620, A5

\bibitem[{{Adami} {et~al.}(2012){Adami}, {Jouvel}, {Guennou}, {Le Brun},
  {Durret}, {Clement}, {Clerc}, {Comer{\'o}n}, {Ilbert}, {Lin}, {Russeil}, \&
  {Seemann}}]{Adami2012}
{Adami}, C., {Jouvel}, S., {Guennou}, L., {et~al.} 2012, \aap, 540, A105

\bibitem[{{Adami} {et~al.}(2007){Adami}, {Russeil}, \& {Durret}}]{Adami2007}
{Adami}, C., {Russeil}, D., \& {Durret}, F. 2007, \aap, 467, 459

\bibitem[{{Aguerri} {et~al.}(2011){Aguerri}, {Girardi}, {Boschin}, {Barrena},
  {M{\'e}ndez-Abreu}, {S{\'a}nchez-Janssen}, {Borgani}, {Castro-Rodriguez},
  {Corsini}, {Del Burgo}, {D'Onghia}, {Iglesias-P{\'a}ramo}, {Napolitano}, \&
  {Vilchez}}]{Aguerri11}
{Aguerri}, J.~A.~L., {Girardi}, M., {Boschin}, W., {et~al.} 2011, \aap, 527,
  A143

\bibitem[{{Alamo-Mart{\'\i}nez} {et~al.}(2012){Alamo-Mart{\'\i}nez}, {West},
  {Blakeslee}, {Gonz{\'a}lez-L{\'o}pezlira}, {Jord{\'a}n}, {Gregg},
  {C{\^o}t{\'e}}, {Drinkwater}, \& {van den Bergh}}]{AlamoMartinez12}
{Alamo-Mart{\'\i}nez}, K.~A., {West}, M.~J., {Blakeslee}, J.~P., {et~al.} 2012,
  \aap, 546, A15

\bibitem[{{Amorisco}(2015)}]{Amorisco15}
{Amorisco}, N.~C. 2015, \mnras, 450, 575

\bibitem[{{Anders} \& {Grevesse}(1989)}]{Anders1989}
{Anders}, E. \& {Grevesse}, N. 1989, \gca, 53, 197

\bibitem[{{Athanassoula} \& {Bosma}(1985)}]{Athanassoula85}
{Athanassoula}, E. \& {Bosma}, A. 1985, \araa, 23, 147

\bibitem[{{Atkinson} {et~al.}(2013){Atkinson}, {Abraham}, \&
  {Ferguson}}]{Atkinson13}
{Atkinson}, A.~M., {Abraham}, R.~G., \& {Ferguson}, A.~M.~N. 2013, \apj, 765,
  28

\bibitem[{{Barrena} {et~al.}(2014){Barrena}, {Girardi}, {Boschin}, {De Grandi},
  \& {Rossetti}}]{Barrena2014}
{Barrena}, R., {Girardi}, M., {Boschin}, W., {De Grandi}, S., \& {Rossetti}, M.
  2014, \mnras, 442, 2216

\bibitem[{{Beers} {et~al.}(1995){Beers}, {Kriessler}, {Bird}, \&
  {Huchra}}]{Beers1995}
{Beers}, T.~C., {Kriessler}, J.~R., {Bird}, C.~M., \& {Huchra}, J.~P. 1995,
  \aj, 109, 874

\bibitem[{{Bender} {et~al.}(1989){Bender}, {Surma}, {Doebereiner},
  {Moellenhoff}, \& {Madejsky}}]{Bender1989}
{Bender}, R., {Surma}, P., {Doebereiner}, S., {Moellenhoff}, C., \& {Madejsky},
  R. 1989, \aap, 217, 35

\bibitem[{{Bharadwaj} {et~al.}(2016){Bharadwaj}, {Reiprich}, {Sanders}, \&
  {Schellenberger}}]{Bharadwaj16}
{Bharadwaj}, V., {Reiprich}, T.~H., {Sanders}, J.~S., \& {Schellenberger}, G.
  2016, \aap, 585, A125

\bibitem[{{B{\'\i}lek} {et~al.}(2016){B{\'\i}lek}, {Cuillandre}, {Gwyn},
  {Ebrov{\'a}}, {Barto{\v{s}}kov{\'a}}, {Jungwiert}, \&
  {J{\'\i}lkov{\'a}}}]{Bilek16}
{B{\'\i}lek}, M., {Cuillandre}, J.~C., {Gwyn}, S., {et~al.} 2016, \aap, 588,
  A77

\bibitem[{{Busko}(1996)}]{Busko1996}
{Busko}, I.~C. 1996, in Astronomical Society of the Pacific Conference Series,
  Vol. 101, Astronomical Data Analysis Software and Systems V, ed. G.~H.
  {Jacoby} \& J.~{Barnes}, 139

\bibitem[{{Canalizo} {et~al.}(2007){Canalizo}, {Bennert}, {Jungwiert},
  {Stockton}, {Schweizer}, {Lacy}, \& {Peng}}]{Canalizo07}
{Canalizo}, G., {Bennert}, N., {Jungwiert}, B., {et~al.} 2007, \apj, 669, 801

\bibitem[{{Cappellari} {et~al.}(2006){Cappellari}, {Bacon}, {Bureau}, {Damen},
  {Davies}, {de Zeeuw}, {Emsellem}, {Falc{\'o}n-Barroso}, {Krajnovi{\'c}},
  {Kuntschner}, {McDermid}, {Peletier}, {Sarzi}, {van den Bosch}, \& {van de
  Ven}}]{Cappellari2006}
{Cappellari}, M., {Bacon}, R., {Bureau}, M., {et~al.} 2006, \mnras, 366, 1126

\bibitem[{{Cooper} {et~al.}(2010){Cooper}, {Cole}, {Frenk}, {White}, {Helly},
  {Benson}, {De Lucia}, {Helmi}, {Jenkins}, {Navarro}, {Springel}, \&
  {Wang}}]{Cooper10}
{Cooper}, A.~P., {Cole}, S., {Frenk}, C.~S., {et~al.} 2010, \mnras, 406, 744

\bibitem[{{Cooper} {et~al.}(2011){Cooper}, {Mart{\'{\i}}nez-Delgado}, {Helly},
  {Frenk}, {Cole}, {Crawford}, {Zibetti}, {Carballo-Bello}, \&
  {GaBany}}]{Cooper11}
{Cooper}, A.~P., {Mart{\'{\i}}nez-Delgado}, D., {Helly}, J., {et~al.} 2011,
  \apjl, 743, L21

\bibitem[{{Dahlem} \& {Thiering}(2000)}]{Dahlem00}
{Dahlem}, M. \& {Thiering}, I. 2000, \pasp, 112, 148

\bibitem[{{Dariush} {et~al.}(2010){Dariush}, {Raychaudhury}, {Ponman},
  {Khosroshahi}, {Benson}, {Bower}, \& {Pearce}}]{Dariush10}
{Dariush}, A.~A., {Raychaudhury}, S., {Ponman}, T.~J., {et~al.} 2010, \mnras,
  405, 1873

\bibitem[{{de Blok} {et~al.}(2014){de Blok}, {J{\'o}zsa}, {Patterson},
  {Gentile}, {Heald}, {J{\"u}tte}, {Kamphuis}, {Rand}, {Serra}, \&
  {Walterbos}}]{deBlok14}
{de Blok}, W.~J.~G., {J{\'o}zsa}, G.~I.~G., {Patterson}, M., {et~al.} 2014,
  \aap, 566, A80

\bibitem[{{Diemand} \& {Moore}(2011)}]{Diemand11}
{Diemand}, J. \& {Moore}, B. 2011, Advanced Science Letters, 4, 297

\bibitem[{{D'Onghia} {et~al.}(2005){D'Onghia}, {Sommer-Larsen}, {Romeo},
  {Burkert}, {Pedersen}, {Portinari}, \& {Rasmussen}}]{Donghia05}
{D'Onghia}, E., {Sommer-Larsen}, J., {Romeo}, A.~D., {et~al.} 2005, \apjl, 630,
  L109

\bibitem[{{Duc}(2016)}]{Duc16}
{Duc}, P.-A. 2016, L'Astronomie, 130, 26

\bibitem[{{Duc} {et~al.}(2011){Duc}, {Cuillandre}, {Serra}, {Michel-Dansac},
  {Ferriere}, {Alatalo}, {Blitz}, {Bois}, {Bournaud}, {Bureau}, {Cappellari},
  {Davies}, {Davis}, {de Zeeuw}, {Emsellem}, {Khochfar}, {Krajnovi{\'c}},
  {Kuntschner}, {Lablanche}, {McDermid}, {Morganti}, {Naab}, {Oosterloo},
  {Sarzi}, {Scott}, {Weijmans}, \& {Young}}]{Duc2011}
{Duc}, P.-A., {Cuillandre}, J.-C., {Serra}, P., {et~al.} 2011, \mnras, 417, 863

\bibitem[{{Duffy} {et~al.}(2008){Duffy}, {Schaye}, {Kay}, \& {Dalla
  Vecchia}}]{Duffy2008}
{Duffy}, A.~R., {Schaye}, J., {Kay}, S.~T., \& {Dalla Vecchia}, C. 2008,
  \mnras, 390, L64

\bibitem[{{D{\"u}nner} {et~al.}(2006){D{\"u}nner}, {Araya}, {Meza}, \&
  {Reisenegger}}]{Dunner06}
{D{\"u}nner}, R., {Araya}, P.~A., {Meza}, A., \& {Reisenegger}, A. 2006,
  \mnras, 366, 803

\bibitem[{{Dupraz} \& {Combes}(1986)}]{Dupraz86}
{Dupraz}, C. \& {Combes}, F. 1986, \aap, 166, 53

\bibitem[{{Dutton} \& {Macci{\`o}}(2014)}]{Dutton2014}
{Dutton}, A.~A. \& {Macci{\`o}}, A.~V. 2014, \mnras, 441, 3359

\bibitem[{{Ebrov{\'a}} {et~al.}(2020){Ebrov{\'a}}, {B{\'\i}lek},
  {Y{\i}ld{\i}z}, \& {Eli{\'a}{\v{s}}ek}}]{Ebrova2020}
{Ebrov{\'a}}, I., {B{\'\i}lek}, M., {Y{\i}ld{\i}z}, M.~K., \&
  {Eli{\'a}{\v{s}}ek}, J. 2020, \aap, 634, A73

\bibitem[{{Foster} {et~al.}(2014){Foster}, {Lux}, {Romanowsky},
  {Mart{\'{\i}}nez-Delgado}, {Zibetti}, {Arnold}, {Brodie}, {Ciardullo},
  {GaBany}, {Merrifield}, {Singh}, \& {Strader}}]{Foster14}
{Foster}, C., {Lux}, H., {Romanowsky}, A.~J., {et~al.} 2014, \mnras, 442, 3544

\bibitem[{{Girardi} {et~al.}(2014){Girardi}, {Aguerri}, {De Grandi},
  {D'Onghia}, {Barrena}, {Boschin}, {M{\'e}ndez-Abreu}, {S{\'a}nchez-Janssen},
  {Zarattini}, {Biviano}, {Castro-Rodriguez}, {Corsini}, {del Burgo},
  {Iglesias-P{\'a}ramo}, \& {Vilchez}}]{Girardi14}
{Girardi}, M., {Aguerri}, J.~A.~L., {De Grandi}, S., {et~al.} 2014, \aap, 565,
  A115

\bibitem[{{Girardi} {et~al.}(2011){Girardi}, {Bardelli}, {Barrena}, {Boschin},
  {Gastaldello}, \& {Nonino}}]{Girardi2011}
{Girardi}, M., {Bardelli}, S., {Barrena}, R., {et~al.} 2011, \aap, 536, A89

\bibitem[{{Guennou} {et~al.}(2014){Guennou}, {Adami}, {Durret}, {Lima Neto},
  {Ulmer}, {Clowe}, {LeBrun}, {Martinet}, {Allam}, {Annis}, {Basa}, {Benoist},
  {Biviano}, {Cappi}, {Cypriano}, {Gavazzi}, {Halliday}, {Ilbert}, {Jullo},
  {Just}, {Limousin}, {M{\'a}rquez}, {Mazure}, {Murphy}, {Plana}, {Rostagni},
  {Russeil}, {Schirmer}, {Slezak}, {Tucker}, {Zaritsky}, \&
  {Ziegler}}]{Guennou2014}
{Guennou}, L., {Adami}, C., {Durret}, F., {et~al.} 2014, \aap, 561, A112

\bibitem[{{Harrison} {et~al.}(2012){Harrison}, {Miller}, {Richards},
  {Lloyd-Davies}, {Hoyle}, {Romer}, {Mehrtens}, {Hilton}, {Stott}, {Capozzi},
  {Collins}, {Deadman}, {Liddle}, {Sahl{\'e}n}, {Stanford}, \&
  {Viana}}]{Harrison2012}
{Harrison}, C.~D., {Miller}, C.~J., {Richards}, J.~W., {et~al.} 2012, \apj,
  752, 12

\bibitem[{{Helmi} {et~al.}(1999){Helmi}, {White}, {de Zeeuw}, \&
  {Zhao}}]{Helmi99}
{Helmi}, A., {White}, S.~D.~M., {de Zeeuw}, P.~T., \& {Zhao}, H. 1999, \nat,
  402, 53

\bibitem[{{Hendel} \& {Johnston}(2015)}]{Hendel15}
{Hendel}, D. \& {Johnston}, K.~V. 2015, \mnras, 454, 2472

\bibitem[{{Hernquist}(1993)}]{Hernquist93}
{Hernquist}, L. 1993, \apjs, 86, 389

\bibitem[{{Jedrzejewski}(1987)}]{Jedrzejewski1987}
{Jedrzejewski}, R.~I. 1987, \mnras, 226, 747

\bibitem[{{Jones} {et~al.}(2003){Jones}, {Ponman}, {Horton}, {Babul},
  {Ebeling}, \& {Burke}}]{Jones03}
{Jones}, L.~R., {Ponman}, T.~J., {Horton}, A., {et~al.} 2003, \mnras, 343, 627

\bibitem[{{Kaastra} \& {Mewe}(1993)}]{Kaastra93}
{Kaastra}, J.~S. \& {Mewe}, R. 1993, \aaps, 97, 443

\bibitem[{{Kanagusuku} {et~al.}(2016){Kanagusuku}, {D{\'{\i}}az-Gim{\'e}nez},
  \& {Zandivarez}}]{Kanagusuku16}
{Kanagusuku}, M.~J., {D{\'{\i}}az-Gim{\'e}nez}, E., \& {Zandivarez}, A. 2016,
  \aap, 586, A40

\bibitem[{{Khosroshahi} {et~al.}(2007){Khosroshahi}, {Ponman}, \&
  {Jones}}]{Khosroshahi2007}
{Khosroshahi}, H.~G., {Ponman}, T.~J., \& {Jones}, L.~R. 2007, \mnras, 377, 595

\bibitem[{{Kim} {et~al.}(2019){Kim}, {Anderson}, {Burke}, {D'Abrusco},
  {Fabbiano}, {Fruscione}, {Lauer}, {McCollough}, {Morgan}, {Mossman},
  {O'Sullivan}, {Paggi}, {Vrtilek}, \& {Trinchieri}}]{Kim19}
{Kim}, D.-W., {Anderson}, C., {Burke}, D., {et~al.} 2019, \apjs, 241, 36

\bibitem[{{Kim} {et~al.}(2018){Kim}, {Anderson}, {Burke}, {Fabbiano},
  {Fruscione}, {Lauer}, {McCollough}, {Morgan}, {Mossman}, \&
  {O'Sullivan}}]{Kim18}
{Kim}, D.-W., {Anderson}, C., {Burke}, D., {et~al.} 2018, \apj, 853, 129

\bibitem[{{Kim} \& {Fabbiano}(2015)}]{Kim15}
{Kim}, D.-W. \& {Fabbiano}, G. 2015, \apj, 812, 127

\bibitem[{{Koranyi} \& {Geller}(2002)}]{Koranyi2002}
{Koranyi}, D.~M. \& {Geller}, M.~J. 2002, \aj, 123, 100

\bibitem[{{Kundert} {et~al.}(2015){Kundert}, {Gastaldello}, {D'Onghia},
  {Girardi}, {Aguerri}, {Barrena}, {Corsini}, {De Grandi},
  {Jim{\'e}nez-Bail{\'o}n}, \& {Lozada-Mu{\~n}oz}}]{Kundert2015}
{Kundert}, A., {Gastaldello}, F., {D'Onghia}, E., {et~al.} 2015, \mnras, 454,
  161

\bibitem[{{La Barbera} {et~al.}(2009){La Barbera}, {de Carvalho}, {de la Rosa},
  {Sorrentino}, {Gal}, \& {Kohl-Moreira}}]{LaBarbera09}
{La Barbera}, F., {de Carvalho}, R.~R., {de la Rosa}, I.~G., {et~al.} 2009,
  \aj, 137, 3942

\bibitem[{{Lagan{\'a}} {et~al.}(2013){Lagan{\'a}}, {Martinet}, {Durret}, {Lima
  Neto}, {Maughan}, \& {Zhang}}]{Lagana2013}
{Lagan{\'a}}, T.~F., {Martinet}, N., {Durret}, F., {et~al.} 2013, \aap, 555,
  A66

\bibitem[{{Lauer}(1985)}]{Lauer1985}
{Lauer}, T.~R. 1985, \mnras, 216, 429

\bibitem[{{Lea}(1975)}]{Lea1975}
{Lea}, S.~M. 1975, \aplett, 16, 141

\bibitem[{{Liedahl} {et~al.}(1995){Liedahl}, {Osterheld}, \&
  {Goldstein}}]{Liedahl95}
{Liedahl}, D.~A., {Osterheld}, A.~L., \& {Goldstein}, W.~H. 1995, \apjl, 438,
  L115

\bibitem[{{Lin} \& {Mohr}(2004)}]{LinMohr04}
{Lin}, Y.-T. \& {Mohr}, J.~J. 2004, \apj, 617, 879

\bibitem[{{Longobardi} {et~al.}(2015){Longobardi}, {Arnaboldi}, {Gerhard}, \&
  {Mihos}}]{Longobardi15}
{Longobardi}, A., {Arnaboldi}, M., {Gerhard}, O., \& {Mihos}, J.~C. 2015, \aap,
  579, L3

\bibitem[{{Malin} \& {Carter}(1980)}]{Malin80}
{Malin}, D.~F. \& {Carter}, D. 1980, \nat, 285, 643

\bibitem[{{Malin} \& {Carter}(1983)}]{Malin83}
{Malin}, D.~F. \& {Carter}, D. 1983, \apj, 274, 534

\bibitem[{{Mart{\'{\i}}nez-Delgado} {et~al.}(2010){Mart{\'{\i}}nez-Delgado},
  {Gabany}, {Crawford}, {Zibetti}, {Majewski}, {Rix}, {Fliri},
  {Carballo-Bello}, {Bardalez-Gagliuffi}, {Pe{\~n}arrubia}, {Chonis}, {Madore},
  {Trujillo}, {Schirmer}, \& {McDavid}}]{Martinez10}
{Mart{\'{\i}}nez-Delgado}, D., {Gabany}, R.~J., {Crawford}, K., {et~al.} 2010,
  \aj, 140, 962

\bibitem[{{Mart{\'{\i}}nez-Delgado} {et~al.}(2012){Mart{\'{\i}}nez-Delgado},
  {Romanowsky}, {Gabany}, {Annibali}, {Arnold}, {Fliri}, {Zibetti}, {van der
  Marel}, {Rix}, {Chonis}, {Carballo-Bello}, {Aloisi}, {Macci{\`o}},
  {Gallego-Laborda}, {Brodie}, \& {Merrifield}}]{Martinez12}
{Mart{\'{\i}}nez-Delgado}, D., {Romanowsky}, A.~J., {Gabany}, R.~J., {et~al.}
  2012, \apjl, 748, L24

\bibitem[{{Mihos} {et~al.}(2017){Mihos}, {Harding}, {Feldmeier}, {Rudick},
  {Janowiecki}, {Morrison}, {Slater}, \& {Watkins}}]{Mihos2017}
{Mihos}, J.~C., {Harding}, P., {Feldmeier}, J.~J., {et~al.} 2017, \apj, 834, 16

\bibitem[{{Moffat}(1969)}]{Moffat69}
{Moffat}, A.~F.~J. 1969, \aap, 3, 455

\bibitem[{{Morgan} {et~al.}(1975){Morgan}, {Kayser}, \& {White}}]{MKW75}
{Morgan}, W.~W., {Kayser}, S., \& {White}, R.~A. 1975, \apj, 199, 545

\bibitem[{{Mulchaey} {et~al.}(1996){Mulchaey}, {Davis}, {Mushotzky}, \&
  {Burstein}}]{Mulchaey96}
{Mulchaey}, J.~S., {Davis}, D.~S., {Mushotzky}, R.~F., \& {Burstein}, D. 1996,
  \apj, 456, 80

\bibitem[{{Navarro} {et~al.}(1997){Navarro}, {Frenk}, \& {White}}]{NFW}
{Navarro}, J.~F., {Frenk}, C.~S., \& {White}, S.~D.~M. 1997, \apj, 490, 493

\bibitem[{{O'Dea} {et~al.}(2008){O'Dea}, {Baum}, {Privon}, {Noel-Storr},
  {Quillen}, {Zufelt}, {Park}, {Edge}, {Russell}, {Fabian}, {Donahue},
  {Sarazin}, {McNamara}, {Bregman}, \& {Egami}}]{ODea08}
{O'Dea}, C.~P., {Baum}, S.~A., {Privon}, G., {et~al.} 2008, \apj, 681, 1035

\bibitem[{{O'Mill} {et~al.}(2015){O'Mill}, {Proust}, {Capelato}, {Castejon},
  {Cypriano}, {Lima Neto}, \& {Laerte}}]{Omill15}
{O'Mill}, A.~L., {Proust}, D., {Capelato}, H.~V., {et~al.} 2015, \mnras, 453,
  868

\bibitem[{{Peng} {et~al.}(2002){Peng}, {Ho}, {Impey}, \& {Rix}}]{Peng02}
{Peng}, C.~Y., {Ho}, L.~C., {Impey}, C.~D., \& {Rix}, H.-W. 2002, \aj, 124, 266

\bibitem[{{Peng} {et~al.}(2010){Peng}, {Ho}, {Impey}, \& {Rix}}]{Peng10}
{Peng}, C.~Y., {Ho}, L.~C., {Impey}, C.~D., \& {Rix}, H.-W. 2010, \aj, 139,
  2097

\bibitem[{{Ponman} {et~al.}(1994){Ponman}, {Allan}, {Jones}, {Merrifield},
  {McHardy}, {Lehto}, \& {Luppino}}]{Ponman1994}
{Ponman}, T.~J., {Allan}, D.~J., {Jones}, L.~R., {et~al.} 1994, \nat, 369, 462

\bibitem[{{Pop} {et~al.}(2018){Pop}, {Pillepich}, {Amorisco}, \&
  {Hernquist}}]{Pop2018}
{Pop}, A.-R., {Pillepich}, A., {Amorisco}, N.~C., \& {Hernquist}, L. 2018,
  \mnras, 480, 1715

\bibitem[{{Quillen} {et~al.}(2008){Quillen}, {Zufelt}, {Park}, {O'Dea}, {Baum},
  {Privon}, {Noel-Storr}, {Edge}, {Russell}, {Fabian}, {Donahue}, {Bregman},
  {McNamara}, \& {Sarazin}}]{Quillen08}
{Quillen}, A.~C., {Zufelt}, N., {Park}, J., {et~al.} 2008, \apjs, 176, 39

\bibitem[{{Quinn}(1984)}]{Quinn84}
{Quinn}, P.~J. 1984, \apj, 279, 596

\bibitem[{{Romanowsky} {et~al.}(2012){Romanowsky}, {Strader}, {Brodie},
  {Mihos}, {Spitler}, {Forbes}, {Foster}, \& {Arnold}}]{Romanowsky12}
{Romanowsky}, A.~J., {Strader}, J., {Brodie}, J.~P., {et~al.} 2012, \apj, 748,
  29

\bibitem[{{Ruszkowski} \& {Springel}(2009)}]{Ruszkowski09}
{Ruszkowski}, M. \& {Springel}, V. 2009, \apj, 696, 1094

\bibitem[{{Santos} {et~al.}(2007){Santos}, {Mendes de Oliveira}, \&
  {Sodr{\'e}}}]{Santos07}
{Santos}, W.~A., {Mendes de Oliveira}, C., \& {Sodr{\'e}}, Laerte, J. 2007,
  \aj, 134, 1551

\bibitem[{{Sarazin}(1986)}]{Sarazin1986}
{Sarazin}, C.~L. 1986, Reviews of Modern Physics, 58, 1

\bibitem[{{Schweizer} \& {Thonnard}(1985)}]{Schweizer85}
{Schweizer}, F. \& {Thonnard}, N. 1985, \pasp, 97, 104

\bibitem[{{Serna} \& {Gerbal}(1996)}]{SernaGerbal96}
{Serna}, A. \& {Gerbal}, D. 1996, \aap, 309, 65

\bibitem[{{Silverman}(1986)}]{Silverman1986}
{Silverman}, B.~W. 1986, {Density Estimation for Statistics and Data Analysis}
  (Chapman and Hall/CRC)

\bibitem[{{Springel}(2005)}]{Springel05}
{Springel}, V. 2005, \mnras, 364, 1105

\bibitem[{{Tal} {et~al.}(2009){Tal}, {van Dokkum}, {Nelan}, \&
  {Bezanson}}]{Tal09}
{Tal}, T., {van Dokkum}, P.~G., {Nelan}, J., \& {Bezanson}, R. 2009, \aj, 138,
  1417

\bibitem[{{Torrey} {et~al.}(2015){Torrey}, {Snyder}, {Vogelsberger}, {Hayward},
  {Genel}, {Sijacki}, {Springel}, {Hernquist}, {Nelson}, {Kriek}, {Pillepich},
  {Sales}, \& {McBride}}]{Torrey15}
{Torrey}, P., {Snyder}, G.~F., {Vogelsberger}, M., {et~al.} 2015, \mnras, 447,
  2753

\bibitem[{{Trujillo} {et~al.}(2001){Trujillo}, {Aguerri}, {Cepa}, \&
  {Guti{\'e}rrez}}]{Trujillo01}
{Trujillo}, I., {Aguerri}, J.~A.~L., {Cepa}, J., \& {Guti{\'e}rrez}, C.~M.
  2001, \mnras, 328, 977

\bibitem[{{Zou} {et~al.}(2016){Zou}, {Maughan}, {Giles}, {Vikhlinin}, {Pacaud},
  {Burenin}, \& {Hornstrup}}]{Zou2016}
{Zou}, S., {Maughan}, B.~J., {Giles}, P.~A., {et~al.} 2016, \mnras, 463, 820

\end{thebibliography}

\end{document}